\documentclass[twocolumn,superscriptaddress,amsfont,amssymb,amsmath, showpacs,balancelastpage, nofootinbib]{revtex4-1}
\usepackage{graphicx,longtable,mathrsfs,color,array}
\usepackage[unicode=true,pdfusetitle,
bookmarks=true,bookmarksnumbered=true,bookmarksopen=true,bookmarksopenlevel=1,
breaklinks=false,pdfborder={0 0 0},backref=false,colorlinks=true]{hyperref}
\hypersetup{citecolor=blue,filecolor=blue,linkcolor=blue,urlcolor=blue}
\usepackage[usenames,dvipsnames]{xcolor}
\usepackage{amssymb,amsmath,mathtools,mathrsfs,enumitem}
\usepackage{epsfig,subfigure,placeins,float}
\usepackage{booktabs,longtable,multirow}
\usepackage{exscale,relsize}
\usepackage[normalem]{ulem}
\usepackage[T1]{fontenc}
\usepackage[utf8]{inputenc}
\usepackage{enumerate}
\usepackage{times, mathptmx}
\usepackage{tikz}
\usepackage{aas_macros}
\usepackage{comment}
\usetikzlibrary{arrows,positioning,decorations.markings,decorations.pathmorphing,calc}
\usepackage[normalem]{ulem}

\usepackage[normalem]{ulem}
\newcommand{\be}{\begin{equation}}
\newcommand{\ee}{\end{equation}}

\newcommand{\aB}{\alpha_B}
\newcommand{\aM}{\alpha_M}

\newcommand{\aK}{\alpha_K}

\newcommand{\gcut}{\gamma_{\rm cut}}

\newcolumntype{C}[1]{>{\centering\let\newline\\\arraybackslash\hspace{0pt}}m{#1}}

\newcommand{\refreplybold}[1]{#1}
\newcommand{\refreplyred}[1]{#1}

\definecolor{hyperref}{RGB}{026,028,087}
\def\gsim{ \lower .75ex \hbox{$\sim$} \llap{\raise .27ex \hbox{$>$}} }
\def\lsim{ \lower .75ex \hbox{$\sim$} \llap{\raise .27ex \hbox{$<$}} }

\newcommand{\dof}{{\it dof} }

\def\nn{\nonumber}
\def\ni{\noindent}

\newcommand*{\mathcolor}{}
\def\mathcolor#1#{\mathcoloraux{#1}}
\newcommand*{\mathcoloraux}[3]{%
\protect\leavevmode
\begingroup
\color#1{#2}#3%
\endgroup
}
\newlength{\stheight}
\newcommand\textst[1][fu-grey]{
\ifmmode\setlength{\stheight}{+1.0ex}
\else\setlength{\stheight}{+0.5ex}
\fi
\bgroup\markoverwith{\textcolor{#1}{\rule[\the\stheight]{2pt}{1.0pt}}}\ULon
} 
\newcommand{\textins}[2][fu-grey]{
\ifmmode\mathcolor{#1}{#2}
\else\textcolor{#1}{#2}\@\,
\fi
}
\graphicspath{{./}}
\allowdisplaybreaks

\tikzstyle{vecArrow} = [thick, decoration={markings,mark=at position
1 with {\arrow[semithick]{open triangle 60}}},
double distance=1.4pt, shorten >= 5.5pt,
preaction = {decorate},
postaction = {draw,line width=1.4pt, white,shorten >= 4.5pt}]

\begin{document}
\title{On tachyonic stability priors for dark energy}

\author{Rafaela Gsponer}
\affiliation{Institute of Cosmology \& Gravitation, University of Portsmouth, Portsmouth, PO1 3FX, U.K.}
\affiliation{Institute for Particle Physics and Astrophysics, ETH Z\"urich, 8093 Z\"urich, Switzerland}

\author{Johannes Noller}
\affiliation{Institute of Cosmology \& Gravitation, University of Portsmouth, Portsmouth, PO1 3FX, U.K.}
\affiliation{Institute for Particle Physics and Astrophysics, ETH Z\"urich, 8093 Z\"urich, Switzerland}
\affiliation{Institute for Theoretical Studies, ETH Z\"urich, Clausiusstrasse 47, 8092 Z\"urich, Switzerland}
\affiliation{DAMTP, University of Cambridge, Wilberforce Road, Cambridge CB3 0WA, U.K.}

\begin{abstract}
A number of stability criteria exist for dark energy theories, associated with requiring the absence of ghost, gradient and tachyonic instabilities. 
Tachyonic instabilities are the least well explored of these in the dark energy context and we here discuss and derive criteria for their presence and size in detail.
Our findings suggest that, while the absence of ghost and gradient instabilities is indeed essential for physically viable models and so priors associated with the absence of such instabilities significantly increase the efficiency of parameter estimations without introducing unphysical biases, this is not the case for tachyonic instabilities. Even strong such instabilities can be present without spoiling the cosmological validity of the underlying models. Therefore, we caution against using exclusion priors based on requiring the absence of (strong) tachyonic instabilities in deriving cosmological parameter constraints.
We illustrate this by explicitly computing such constraints within the context of Horndeski theories, while quantifying the size and effect of related tachyonic instabilities.  
\end{abstract}

\date{\today}
\maketitle

\section{Introduction} \label{sec-intro}

Theoretical priors are an essential ingredient for the efficient computation of cosmological dark energy constraints. Their usefulness can come in two flavours: They can increase the efficiency of the computation, by allowing one to {\it a priori} exclude regions of parameter space, which the data would have otherwise excluded by themselves {\it a posteriori}, {and/or} they allow us to take into account information from additional physical requirements that the computation would otherwise not have been directly sensitive to.
With priors of the first kind we are improving the efficiency of constraint extraction, while in the second case we are using complementary physical insights to inform our understanding of (in the present case) cosmological dark energy.  

Here we will primarily be concerned with theoretical priors of the first kind. In the context of dynamical dark energy theories, stability criteria associated with classical background and linear perturbative evolutions are perhaps the most straightforward example of such priors. 
Linear fluctuations on top of a cosmological background can in principle display three types of such instabilities: Ghost, gradient and tachyonic. 
The first two are the better understood, with any such instability generically invalidating the associated theory. 
As such, requiring their absence is frequently incorporated into standard cosmological parameter analyses as a theoretical prior, e.g. as part of the {\it hi\_class} \cite{Zumalacarregui:2016pph} and {\it EFTCAMB} \cite{Hu:2013twa} Einstein-Boltzmann solvers -- see \cite{mcmc,BelliniParam,Hu:2013twa,Raveri:2014cka,Gleyzes:2015rua,Kreisch:2017uet,Zumalacarregui:2016pph,Alonso:2016suf,Arai:2017hxj,Frusciante:2018jzw,Reischke:2018ooh,Mancini:2018qtb,Brando:2019xbv,Arjona:2019rfn,Raveri:2019mxg,Perenon:2019dpc,Frusciante:2019xia,SpurioMancini:2019rxy,Bonilla:2019mbm,Baker:2020apq,Joudaki:2020shz,Noller:2020lav,Noller:2020afd} as examples of recent related cosmological parameter constraints and forecasts in the dark energy context relevant to this paper.
As we will discuss, since the presence of such instabilities generically signals unphysical theories, this is typically a safe procedure, which saves computational time (theories displaying such instabilities would otherwise mostly be ruled out by the data {\it a posteriori}) while only introducing minimal errors (associated to non-generic cases -- see e.g. related discussions in \cite{mcmc,radstab}). 

Tachyonic instabilities instead are not as clear-cut and in fact their presence can be observationally required  -- the Jeans' instability in standard $\Lambda{}$CDM cosmology can be viewed as such a (tachyonic) instability. 
However, it has previously been conjectured (see e.g. discussions in \cite{DeFelice:2016ucp,Lagos:2017hdr,Frusciante:2018vht}) that only mild such instabilities, which evolve on cosmological time-scales similar to those associated with the Jeans' instability, lead to observationally viable theories. Stronger tachyonic instabilities would then be associated to unphysical theories. So upon accurately identifying the transition scale between such mild and strong instabilities, requiring the absence of strong tachyonic instabilities could be used as a theoretical prior just as in the case of ghost and gradient instabilities. In this paper we explore whether this is possible, i.e. whether a meaningful transition scale between mild and strong tachyonic instabilities exists, which can be used to place a theoretical prior on cosmological parameter estimation that improves its efficiency.

\section{The setup} \label{sec-setup}

\ni {\bf Horndeski gravity}:
We will be working within the context of Horndeski gravity~\cite{Horndeski:1974wa,Deffayet:2011gz},\footnote{For the equivalence between the formulations of \cite{Horndeski:1974wa} and \cite{Deffayet:2011gz}, see \cite{Kobayashi:2011nu}.}
the most general scalar-tensor theory, which gives rise to second order equations of motion for the metric $g_{\mu\nu}$ and the additional scalar field $\phi$. As such, we are considering theories where dark energy is described by a single additional gravitational degree of freedom ({\it dof}). More specifically, we will consider theories with the following Lagrangian
\be
{\cal L} = G_2(\phi, X) - G_3(\phi, X) \square \phi + G_{4}(\phi) R + \mathcal{L}_m,
    \label{Horndeski}
\ee
where the $G_i$ are free functions of the scalar field $\phi$ and its derivative via $X = - \frac{1}{2} \nabla^{ \mu} \phi \nabla_{\mu} \phi$ and $\mathcal{L}_m$ describes the matter sector for all matter fields,
which is minimally coupled to gravity and hence independent of $\phi$ (in other words, here we are working in Jordan frame). Note that \eqref{Horndeski} is the subset of theories that give rise to gravitational waves propagating precisely at the speed of light \cite{Creminelli:2017sry, Sakstein:2017xjx, Ezquiaga:2017ekz, Baker:2017hug} -- also see \cite{Amendola:2012ky,Amendola:2014wma,Deffayet:2010qz,Linder:2014fna,Raveri:2014eea,Saltas:2014dha,Lombriser:2016yzn,Lombriser:2015sxa,Jimenez:2015bwa,Bettoni:2016mij,Bettoni:2016mij,Sawicki:2016klv} for closely related prior work.
\\

\ni {\bf Modelling matter}: The matter part of the Lagrangian, $\mathcal{L}_m$ in \eqref{Horndeski}, in principle contains several degrees of freedom, all of which come with their own individual stability criteria. In practice, however, these are not modelled individually and so we will follow a hybrid approach here. At the level of the cosmological constraint analyses we will present later in this paper, we will model matter as a mixture of two effective fluids describing non-relativistic matter and radiation, respectively. This is the standard approach implemented in the aforementioned Einstein-Boltzmann solvers. However, for the analytic derivation of stability priors we will follow the methodology outlined in \cite{Lagos:2017hdr} and derive stability conditions by working with the following matter scalar Lagrangian
\begin{align}
{\mathcal L}_m = - \frac{1}{2} \partial_{\mu} \chi \partial^{\mu}\chi - V(\chi),
\label{Lmatter}
\end{align}
where $\chi$ is a canonical matter scalar and $V$ an arbitrary potential. Clearly the matter proxy \eqref{Lmatter} cannot accurately mimic the full complexity of a cosmological matter fluid and its components, but following \cite{Lagos:2017hdr} we will assume that the stability conditions derived with \eqref{Lmatter} are valid for a general matter fluid, with the expectation that additional stability conditions would arise when more fully modelling all matter degrees of freedom. The conditions derived here would then be conservative tracers of the full physical set of stability conditions.  We refer to section \ref{sec-Jeans} for a more detailed discussion of how different matter models and components can lead to different (complementary) stabilty conditions.
\\

 \begin{figure}[t!]
\centering
\includegraphics[width=0.9\linewidth]{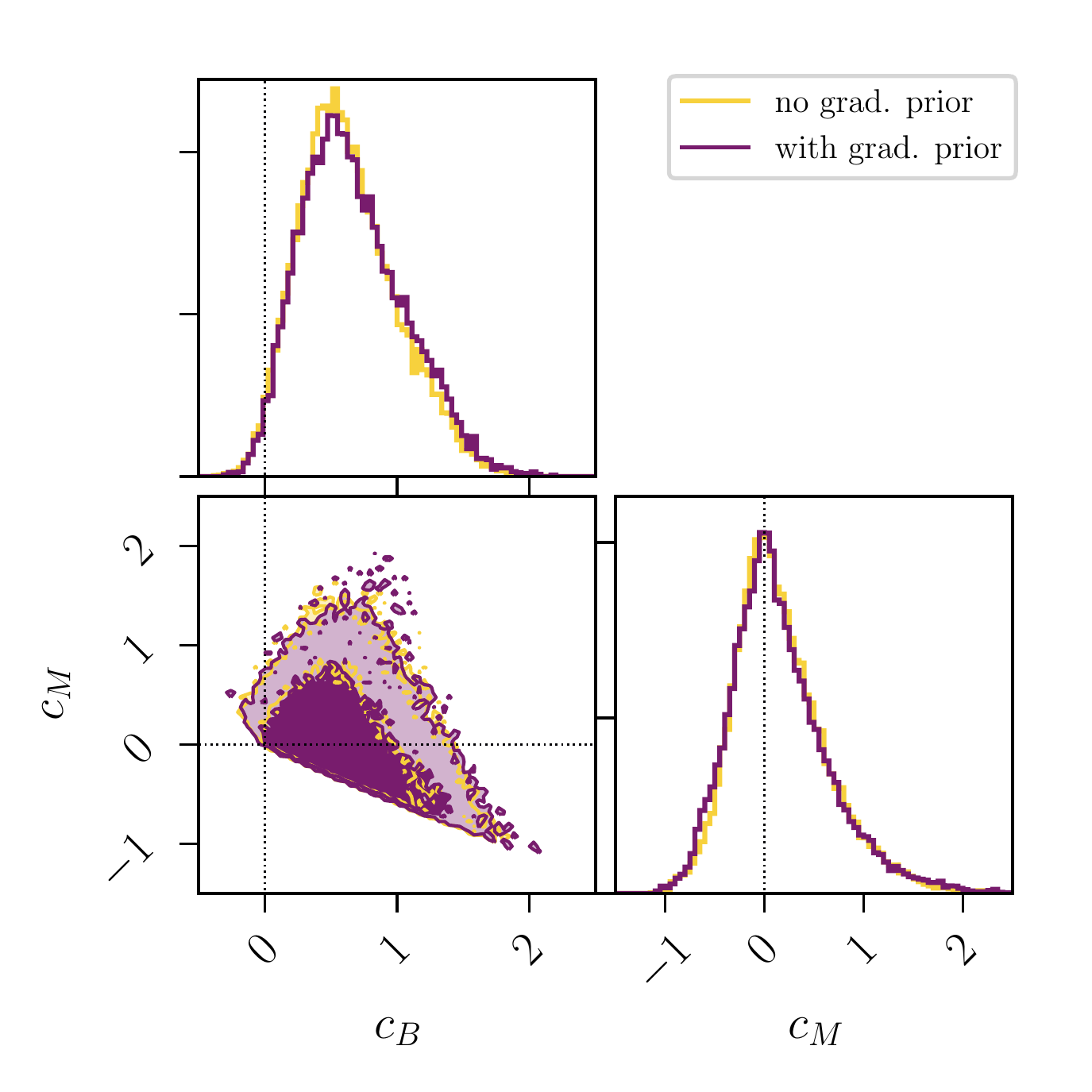}
\caption{
Cosmological parameter constraints for the modified gravity parameters $c_b$ and
$c_m$, using the parameterisation~(\ref{param1}) and CMB, RSD, BAO and mPk measurements (see section \ref{sec-setup} for details).
Inner (outer) contours correspond to $68 \% \; ( 95 \% )$ confidence levels.
Here we contrast constraints obtained without vs. with gradient stability priors (without tachyonic stability priors in both cases). The constraints are near identical, although note that even these very minor differences can matter for specific well-motivated models (see main text for discussion).}
\label{gradientcut}
\end{figure}

\ni {\bf Linear cosmology}: 
In analysing and constraining cosmological deviations from GR, we will focus on linear perturbations around a $\Lambda{}$CDM background. 
The background equations then read:
\be 
H^2 = \rho_{tot}, \quad \dot{H} = - \frac{3}{2}(\rho_{tot} + p_{tot}),
\label{background}
\ee
where $\rho_{tot}$ and $p_{tot}$ are the total energy density and pressure in the Universe, respectively.\footnote{Note that we are using CLASS units, setting $8 \pi G = 1$. Also there is a relative factor of 3 in the definitions of densities and pressures used here, when compared to other frequently used conventions.}  
The freedom of the linear perturbations of \eqref{Horndeski} around this $\Lambda{}$CDM background are then controlled by three functions $\alpha_i$ given by~\cite{Bellini:2014fua}:
\be
\begin{aligned}
    H M^2 \alpha_M = &\frac{d}{dt}M^2 = 2\dot\phi G_{4 \phi} ,\\
        H^2 M^2 \alpha_K = &2X ( G_{2X} + 2 X G_{2XX} \\
        &- 2G_{3\phi} - 2X G_{3 \phi X})\\ &+12 \dot{\phi} X H ( G_{3 X} + X G_{3XX}) ,\\
        H M^2 \alpha_B = &2 \dot{\phi}( XG_{3X} - G_{4 \phi}),
\end{aligned}
\label{alphas}
\ee
where $M$ is the effective Planck mass and satisfies $M^2 = 2 G_4$. The subscript $X$ represents a partial derivative w.r.t $X$, while a subscript $\phi$ denotes a partial derivative w.r.t the field $\phi$. Here the \textit{kineticity} $\alpha_K$ describes a contribution towards the kinetic energy term of the scalar perturbations, the \textit{braiding} $\alpha_B$ signifies the mixing of the kinetic terms of the scalar and tensor perturbations and the \textit{Planck-mass run rate} $\alpha_M$ quantifies the rate of evolution of the effective Planck mass $M$. 

\begin{figure}[t!]
\centering
\includegraphics[width=0.9\linewidth]{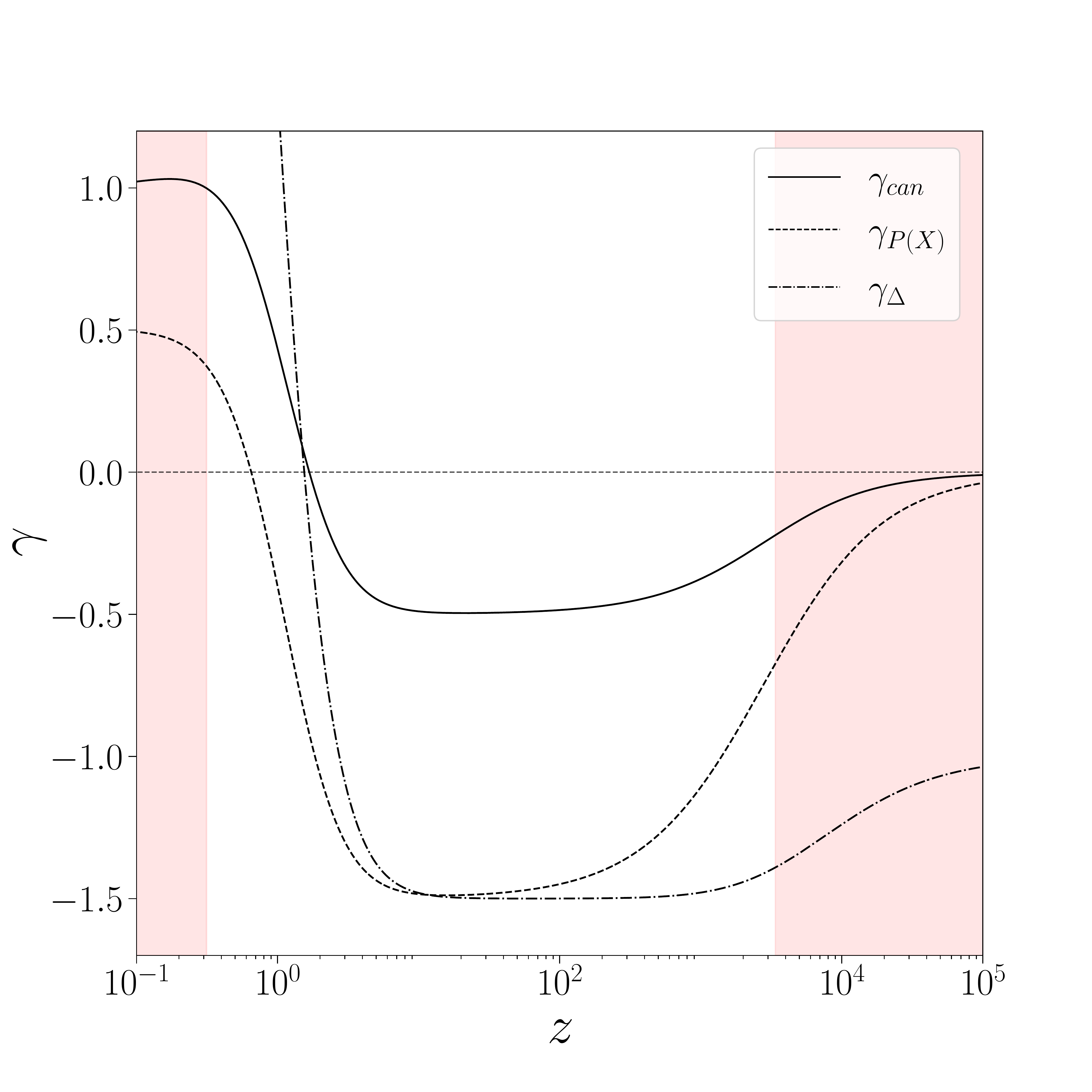}
\caption{ Evolution of the tachyonic instability parameter $\gamma$ vs. redshift $z$ in standard $\Lambda{}$CDM as derived using two different matter proxies. 
$\gamma_{can}$ (solid line) is derived using a canonical scalar field as a matter proxy, corresponding to Eq.~\eqref{JM}, whereas $\gamma_{P(X)}$ (dashed line) is derived using a k-essence-like $P(X)$ scalar as a matter proxy, corresponding to Eq.~\eqref{JMcomp}.
For the example shown we see that the instability condition derived using a canonical scalar matter proxy is always weaker than that using the more involved $P(X)$ matter proxy, so $\gamma_{can}$ here acts as a conservative tracer of the overall instability conditions. 
In an abuse of notation, we also define and plot $\gamma_\Delta$ for comparison (dash-dotted line), which is defined to be the coefficient of $\Delta$ in \eqref{DeltaEqn} divided by $H^2$. In other words, $\gamma_\Delta$ quantifies the strength of the standard Jeans' instability. The plot then emphasises that the above three conditions are not identical (though a precise mapping can be established) -- see section \ref{sec-tach} for a more detailed discussion.
The shaded regions correspond to radiation (right) and dark energy (left) dominated eras, respectively.
}
\label{JeansMass}
\end{figure}

In order to constrain the freedom inherent in the $\alpha_i$ functions, it is useful to choose a parametrisation for them. 
In this paper, we will primarily show results for one of the most commonly used such parameterisations
\begin{equation}
    \alpha_i = c_i \Omega_{DE},
    \label{param1}
\end{equation}
a one parameter ansatz where all the $\alpha_i$ are proportional to the fractional density of the dark energy fluid at the background level. For comparison we will discuss results for another commonly used ansatz, namely, $\alpha_i = c_i a$, in appendix~\ref{appendix_param}.
See \cite{Bellini:2014fua,BelliniParam,Linder:2015rcz,Linder:2016wqw,Denissenya:2018mqs,Lombriser:2018olq,Gleyzes:2017kpi,Alonso:2016suf,mcmc,Noller:2020afd} and especially \cite{Traykova:2021hbr} for a more detailed discussion of the relative merits of different such parametrisations.
Both of the above-mentioned parameterisations ensure that any modifications to GR phenomenology only become important once dark energy plays a significant role at the background level, i.e. at late times.\\

\begin{figure*}[t!]
\includegraphics[width=.45\linewidth]{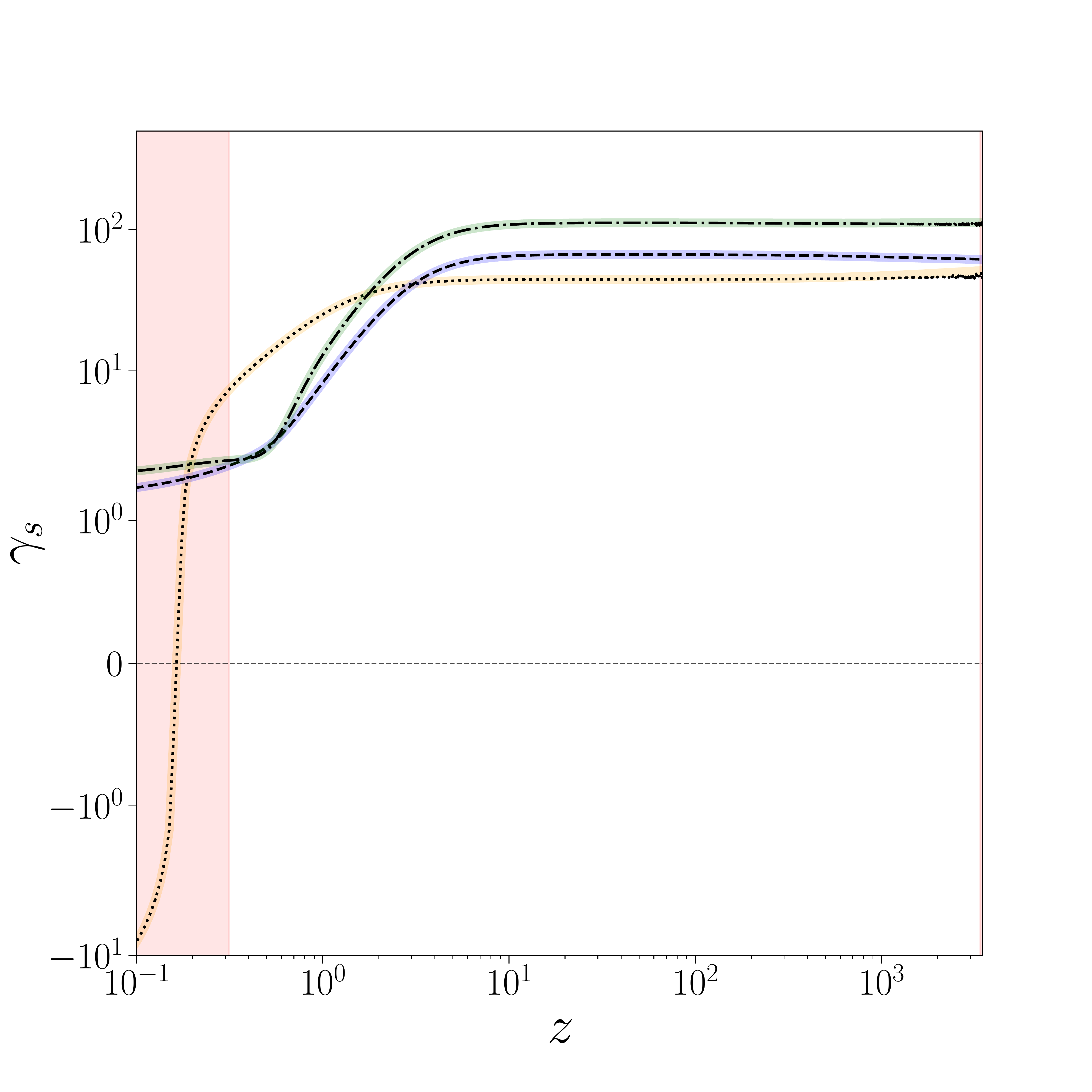}
\includegraphics[width=.45\linewidth]{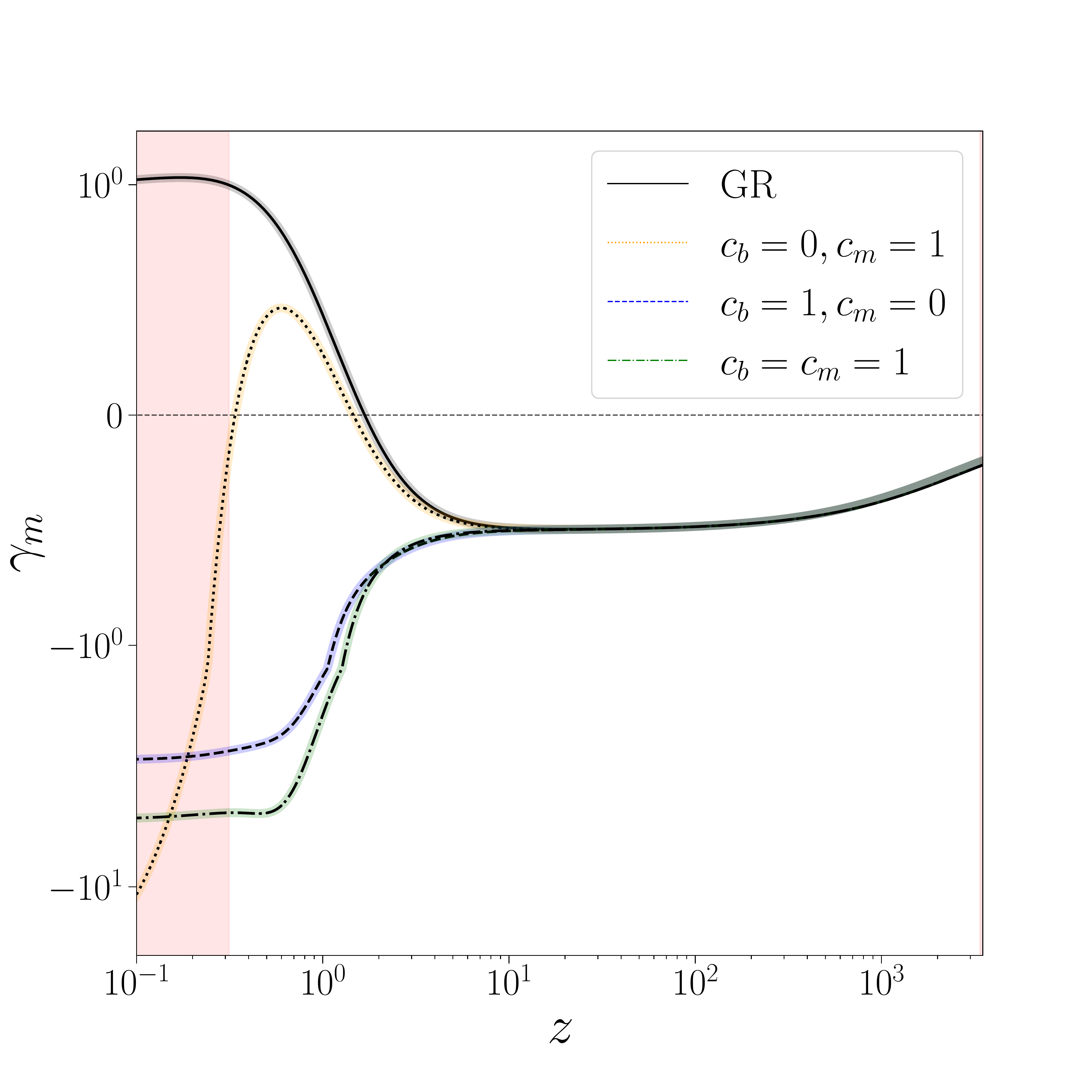}
\caption{
Evolution of $\gamma_s$ and $\gamma_m$, where $c_k$ is fixed to 0.1 and $c_b$ and $c_m$ are varied according to the legend, all other cosmological
parameters were set to their Planck best fit value~\cite{Ade:2015xua}. Note that our  {\it hi\_class}-based implementation (fainter,broader lines) is in perfect agreement with the independent analytical cross-checks (solid/dotted/dashed/dashed-dotted lines). The shaded region on the left corresponds to the dark energy dominated era.
}
\label{fig-gammas}
\end{figure*}

\ni {\bf Cosmological constraints}: 
We will compute cosmological parameter constraints on the dark energy/modified gravity $c_i$ parameters via a Markov chain Monte Carlo (MCMC) analysis,
while marginalising over the standard $\Lambda{\rm CDM}$ parameters $\Omega_{\rm cdm}, \Omega_{\rm b}, \theta_s,A_s,n_s$ and $\tau_{\rm reio}$.
The data sets used are: Planck 2015 CMB temperature, CMB lensing and low-l polarisation data~\cite{Planck-Collaboration:2016aa,Planck-Collaboration:2016af,Planck-Collaboration:2016ae},\footnote{\refreplybold{We note that the most significant difference between the recent Planck 2018 results~\cite{Planck:2019nip} and the earlier results from 2015 used here is the shift to a lower value of the optical depth to reionization, $\tau_{\mathrm{reion}}$, by approximately $1.5 \sigma$. As there are no strong correlations between the value of $\tau_{\mathrm{reion}}$ and the $\alpha_i$ parameters, we believe that these new constraints will not significantly affect our conclusions} \refreplyred{ -- see \cite{mcmc} for a related discussion.} \refreplyred{However, more explicitly analysing any constraint changes induced by using more recent and/or additional CMB data will be an interesting task left for future work here, especially when going beyond cosmologies with $\Lambda{}$CDM backgrounds as considered here.}} baryonic acoustic oscillation (BAO) measurements from SDSS/BOSS~\cite{Anderson:2014,Ross:2015}, data from the SDSS DR4 LRG matter power spectrum shape~\cite{Tegmark:2006} and redshift space distortion (RSD) measurements from BOSS and 6dF~\cite{Beutler:2012,Samushia:2014}.
For technical details regarding the MCMC implementation and for a discussion of related cosmological parameter constraints (as well as for additional details on the implementation and use of the data sets involved) see \cite{mcmc}. 
Finally note that, while we will see that $\aK$ does impact the form taken by tachyonic instabilities, it is well known that it is almost unconstrained by  observations of cosmological linear perturbations~\cite{BelliniParam,Alonso:2016suf}, which is linked to the fact that it is not present in the equation of motion in the quasi-static approximation \cite{Bellini:2014fua}. \refreplybold{Mimicking the above-mentioned previous analyses, we will therefore mostly fix a fiducial behaviour for $\alpha_K$ by setting $c_K = 0.1$ -- see appendix~\ref{appendix_aK} for a detailed discussion of the effect $\alpha_K$ (and hence setting a given fiducial value of $c_K$) has on tachyonic instabilities.}
\\

\section{Ghost and gradient stability criteria} \label{sec-grad}
Of the three instabilities we will discuss, ghost and gradient instabilities are the most well understood. Ghosts are associated with a negative kinetic energy term (and, in addition to triggering classical instabilities, also disastrous when taking into account quantum fluctuations), gradient instabilities occur when the `sound speed' of fluctuations turns imaginary (generically resulting in an uncontrollable growth of perturbations) and tachyonic instabilities are associated with an imaginary effective mass for the fluctuations. 
For illustration, consider the following Lagrangian for a (e.g. matter or Horndeski scalar) fluctuation $\pi$ on top of a purely time-dependent FRW background 
\begin{align}
{{\cal S}^{(2)}} &= \int d^3 k d\tau \big [\pm (\pi^\prime)^2 - (c_s^2 k^2 + \tilde{\mu}^2) \pi^2 \big ],
\label{action1SF}
\end{align}
where $\tau$ is conformal time, a prime denotes a derivative with respect to $\tau$ and we have canonically normalised the scalar $\pi$ up to an overall sign here. Switching to physical time $t$ this action then becomes 
\be 
{{\cal S}^{(2)}} = \int d^3 k dt a \big [\pm (\dot \pi)^2 - (\tfrac{c_s^2 k^2}{a^2} + \mu^2) \pi^2 \big ],
\label{stab_schematic}
\ee
where we have absorbed a factor of $a$ into the re-defined effective mass parameter $\mu^2$. If the first term carries a negative sign, $-(\dot\pi)^2$, then a ghost instability is present, while a gradient instability occurs when $c_s^2 < 0$. A tachyonic instability is present whenever $\mu^2 < 0$.
We will now first summarise when such ghost and gradient instabilities occur for \eqref{Horndeski}, before discussing conditions for the presence of tachyonic instabilities in the next section.
\\

\begin{figure*}[t!]
\includegraphics[width=\linewidth]{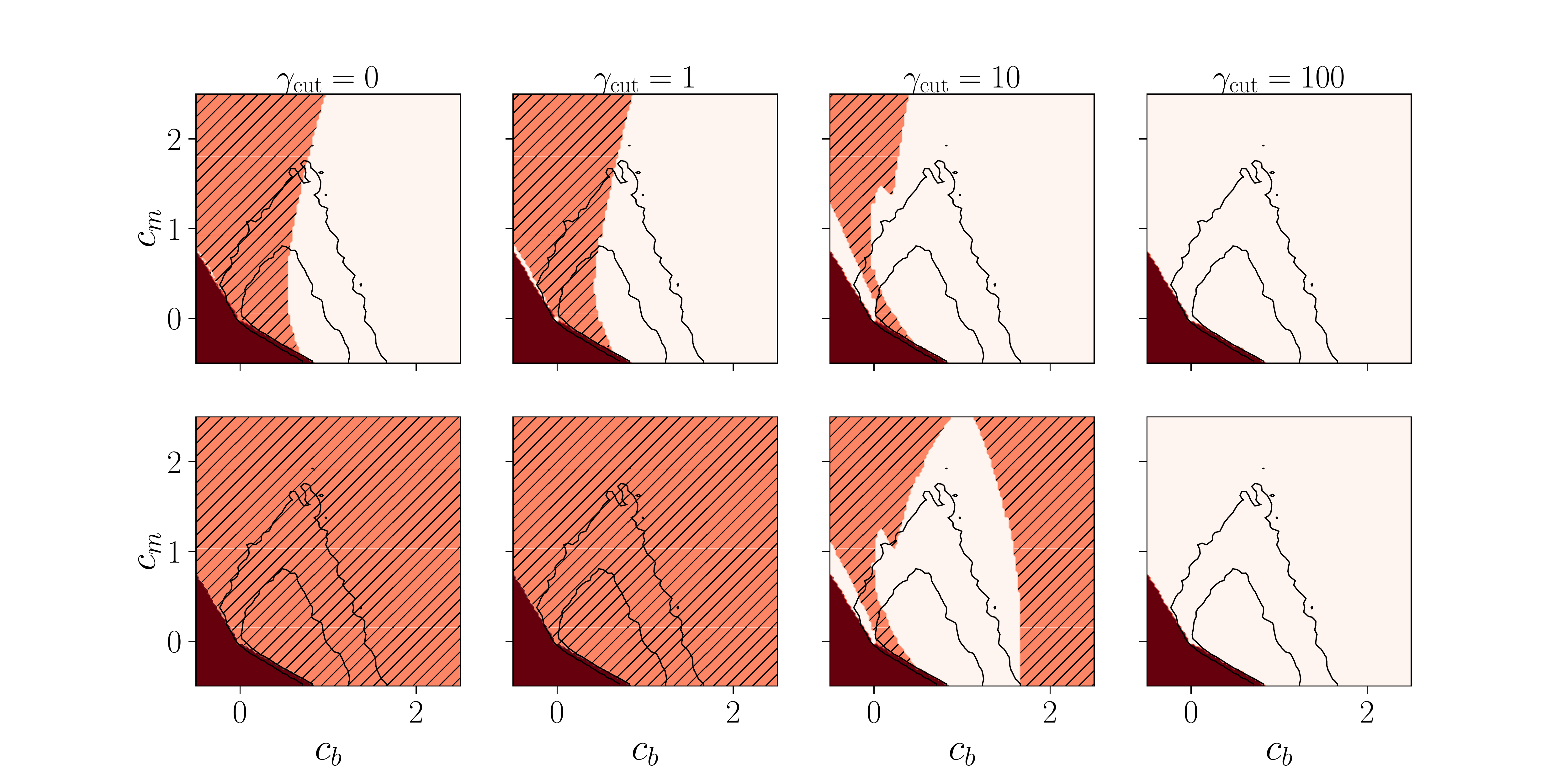}
\caption{
Here we show region plots of the parameter space for $c_m$ and $c_b$ divided in regions where gradient (red), tachyonic (orange-striped) instabilities or no/soft instabilities (white) are present for different values of $\gamma_{\rm cut}$. The black contours corresponds to cosmological parameter constraints generated without tachyonic stability priors as shown in fig. \ref{gradientcut}, with inner (outer) contours corresponding to $68 \% \; ( 95 \% )$ confidence levels.
\textbf{Upper row:} Here $\gamma_{\rm cut}$ is only applied to $\gamma_s$, i.e. to the dark energy scalar sector. \textbf{Lower row:} Here the scalar \textit{and} the matter tachyonic stability condition are applied, thus increasing the amount of parameter space identified as being affected by tachyonic instabilities. We see that for $\gamma_{\rm cut} = 1$ nearly the whole parameter space  experiences a tachyonic instability. Note that this is not surprising, since we expect that observationally viable theories can and do experience (at least soft) tachyonic instabilities of the order of the Jeans instability (i.e. with $\gamma_m \sim {\cal O}(1)$) in the matter sector.
Finally note that the origin does not exactly correspond to $\Lambda{}$CDM here, since we fix a fiducial $c_k = 0.1$ here.
\label{fig-tachyRegion}}
\end{figure*}

\ni {\bf Ghost instabilities}: 
For \eqref{Horndeski}, requiring the absence of ghost instabilities for scalar fluctuations amounts to
\be \label{ghost_cond}
\mathcal{D} \equiv \frac{3}{2} \alpha_B^2 + \alpha_K > 0.
\ee
This condition depends on both $\alpha_B$ and $\alpha_K$, so does indeed imply an implicit constraint on $\alpha_K$ in terms of $\alpha_B$.  
The aforementioned statement that $\alpha_K$ is (a combination of the $G_i$ and their derivatives that is) effectively ’orthogonal’ to the parameter space probed by linear cosmology-related observations is to be understood as applying after this implicit constraint is in place.
Secondly, notice that the above condition is $k$-independent, so if such an instability is present on some fiducial scale $k_{\rm fid}$, it will be present at all scales (i.e. there is no sense in which such a ghost can be regulated as a small-$k$ ghost \cite{Gumrukcuoglu:2016jbh,Lagos:2017hdr,Wolf:2019hzy}).
\\

\ni {\bf Gradient instabilities}: 
The speed of sound for scalar perturbations from \eqref{Horndeski} satisfies
\begin{align} \label{sound_speed}
{\cal D} c_s^2 = (2-\aB)\left(\frac{\aB}{2} + \aM - \frac{\dot H}{H^2}\right) - \frac{3(\rho_{\rm tot} + p_{\rm tot})}{H^2M^2} + \frac{\dot{\alpha}_B}{H},
\end{align}
where we recall that ${\cal D} $ is positive by virtue of requiring the absence of ghosts. As discussed above, requiring the absence of gradient instabilities then amounts to $c_s^2 \geq 0$.
In fig.~(\ref{gradientcut}) we show the effect of extracting cosmological parameter constraints with vs. without imposing this requirement as a prior. One can clearly see that there is excellent overlap between the two regions, signalling that the data by themselves exclude the vast majority of parameter space regions {\it a posteriori}, if they haven't already been excluded by priors in the analysis. Imposing a gradient stability prior therefore appears well-motivated, significantly increasing the efficiency of parameter space estimation (in practice, unstable parameter space regions take significantly longer to compute than stable ones) without introducing unphysical artefacts in the eventual constraints.\footnote{Note, however, that this statement should be interpreted cautiously. Interesting, albeit small, regions of parameter space may still (erroneously) be excluded in this way -- see \cite{radstab,Noller:2020afd} for a discussion of such cases and how to remedy these issues.}
For a discussion of related results see \cite{mcmc}.

\begin{figure*}[t!]
\includegraphics[width=0.45\linewidth]{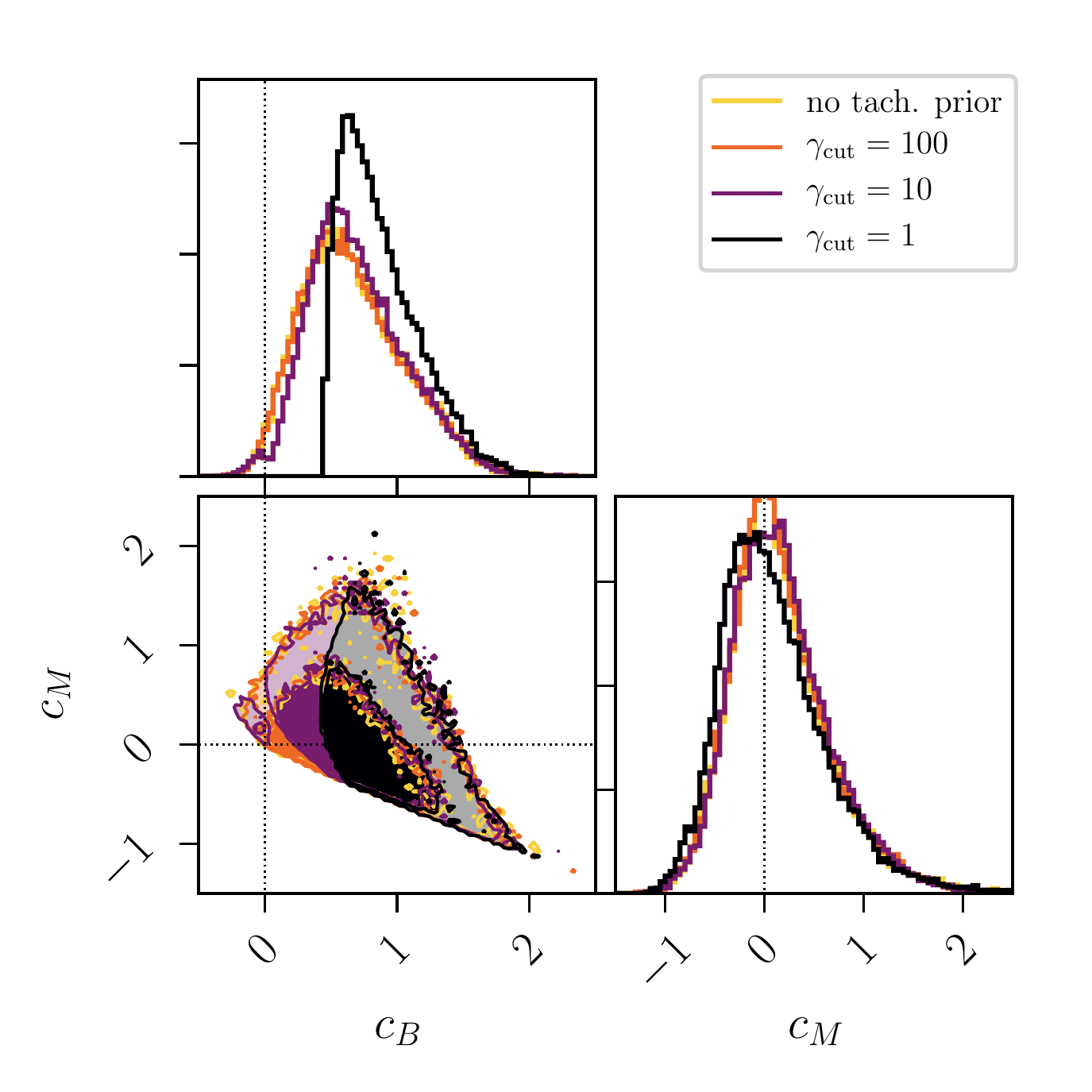}
\includegraphics[width=0.45\linewidth]{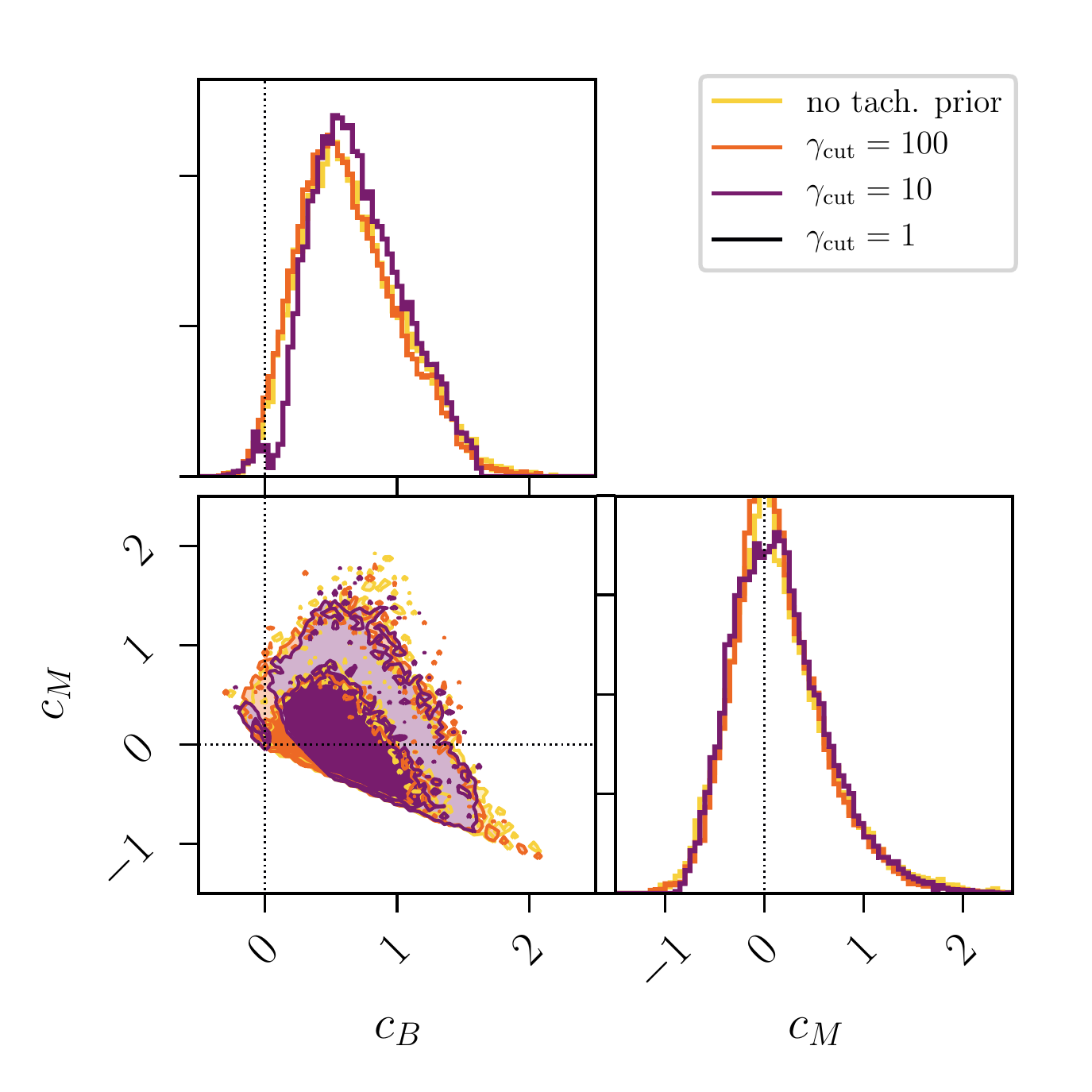}
\caption{
Cosmological parameter constraints for the modified gravity parameters $c_b$ and
$c_m$, using the parameterisation~(\ref{param1}).  Inner (outer) contours correspond to $68 \% ( 95 \% )$ confidence levels.  {\bf Left}: Here we show how constraints change when cuts are imposed on $\gamma_{\rm cut}$ for the scalar tachyonic stability condition $\gamma_s$ only. The full observationally viable region is only recovered for $\gamma_{\rm cut} \gtrsim 100$, as expected from figure \ref{fig-tachyRegion}. However, for such a large $\gamma_{\rm cut}$ the prior also does not exclude any of the relevant non-viable regions of parameter space, so setting this prior then effectively has no effect at all.  {\bf Right}: Here we show how constraints change when cuts are imposed on $\gamma_{\rm cut}$ for the scalar {\it and} for the matter tachyonic stability condition, i.e. for both $\gamma_s$ and $\gamma_m$. Again as expected from figure \ref{fig-tachyRegion}, including a tachyonic stability prior for the matter mode excludes all cosmologically viable models when setting $\gamma_{\rm cut} \lesssim 1$ (so no contours are visible for this value), but only induces fairly minimal changes when $\gamma_{\rm cut} \gtrsim 10$.
\label{fig-tachyConstraints}}
\end{figure*}

\section{Tachyonic instabilities in GR} \label{sec-Jeans}

Before considering the full Horndeski case, we would like to build some intuition by considering tachyonic instabilities in standard General Relativity, here specifically for $\Lambda{}$CDM cosmologies. In such cosmologies gravitational collapse is associated to the well-known Jeans instability~\cite{1902RSPTA.199....1J}, which can be re-cast as a tachyonic instability. 
Here we will use the GR/$\Lambda{}$CDM example to elucidate the link and mapping between a tachyonic instability such as the Jeans instability and the stability properties derived in this paper using \eqref{Lmatter}. 
To do so, we will first look at the standard Jeans instability in an effective fluid picture for matter and then discuss how such a fluid Jeans instability is related to a tachyonic instability of an effective matter scalar \eqref{Lmatter}.
\\

\ni {\bf Fluid picture}:  We begin by modelling matter via the following fluid stress-energy tensor
\be
T_{\mu \nu}^{fluid} = g_{\mu \nu} p_m + ( \rho_m + p_m ) u_{\mu} u_{\nu} + \sigma_{\mu \nu},
\label{Tfluid}
\ee
where $\sigma$ is the anisotropic stress and $u$ is the 4-velocity defined as $u = (1/a (1-\Psi), 1/a^2\delta^{ij} \partial_j v_m))$. In addition we will work in an FRW space-time and in Newtonian Gauge for the associated perturbations, where the line element can be written as:
\be
ds^2 =  - (1 + 2 \Psi ) d t^2  + a^2 (1 - 2 \Phi) \delta_{ij} dx^i dx^j .
\ee
The background Friedmann equations \eqref{background} as well as the continuity equation $\dot{\rho}_m = - 3 H (\rho_m + p_m)$ can then be obtained from the background Einstein and stress-energy conservation equations. 
For the purpose of this paper we set the anisotropic stress $\sigma$ to zero, a property that is recovered by the scalar field matter proxies we will encounter below.

Linearly perturbing the Einstein equations and considering the individual components of the resulting $\delta G_{\mu\nu} = \delta T_{\mu\nu}$, the tracefree part of the $ij$ component of the perturbed Einstein Equation then implies that $\Psi = \Phi$ (contingent on the above $\sigma = 0$ assumption). 
Combining the $00$ and the $0i$ components of this equation, we obtain the Poisson equation
\be
 \nabla^2 \Psi = \frac{3}{2} a^2 \rho_m \Delta,
 \label{Poisson}
\ee
where $\Delta \equiv \delta_m + (\dot{\rho} v_m)/\rho$ is defined to be the gauge-invariant comoving density perturbation, here given in terms of the fractional overdensity $\delta_m = \delta \rho_m /\rho_m$ and the peculiar velocity $v_m$.
Complementing the perturbed Einstein equations we also have the perturbed stress-energy conservation equation, i.e.  $\delta \left[\nabla^{\mu} T_{\mu \nu}\right] = 0$, essentially the equations of motion of the fluid perturbations. We find
\begin{align}
   \delta \dot{\rho}_m +3 H (\delta \rho_m + \delta p_m)+ \frac{(\rho_m + p_m)}{a^2} \nabla^2 v_m &= 3 \dot{\Psi} (\rho_m + p_m), \nonumber \\
     \nabla^2 \dot{v}_m + \nabla^2 \Psi + \frac{\nabla^2 \delta p_m}{(\rho_m+p_m)}+ \frac{\dot{p}_m \nabla^2 v_m}{\rho_m + p_m} &= 0,
\label{eqn-pertContEuler}
\end{align}
so effectively the perturbed continuity and Euler equation.
Combining (the time derivative of) the Poisson Equation~(\ref{Poisson}) with the gradient of the $ii$ component of the perturbed Einstein Equation and \eqref{eqn-pertContEuler}, we then obtain
\be
\ddot{\Delta} + (4 \frac{\dot{H}}{H}+ 8 H) \dot{\Delta} +(20 \dot{H} + \frac{2 \dot{H}^2}{H^2}+ \frac{2 \ddot{H}}{H} +15 H^2) \Delta = \frac{\nabla^2 \delta p_m}{ H^2 a^2},
\label{DeltaEqn}
\ee
where we have made use of a quasi-static approximation, here specifically $\tfrac{\rm d}{\rm dt}{\Psi} \sim 0$.
In matter domination, where $\dot{H} = -\tfrac{3}{2} H^2$ and pressure fluctuations are zero,
this then finally reduces to the well-known evolution equation for non-relativistic matter fluctuations $\Delta$ 
\be 
\ddot{\Delta}_{\rm m} + 2 H \dot{\Delta}_{\rm m} - \frac{3}{2} H^2 \Delta_{\rm m} = 0.
\label{Delta}
\ee
Eq.~\eqref{Delta} neatly illustrates the Jeans instability as encoded in (the sign of) its final term. This instability here is of order $H^2$, so as one would expect, the Hubble scale $H$ is the time scale associated to this instability in $\Lambda{}$CDM. As discussed above, the Jeans instability can be re-cast as a tachyonic instability.\footnote{An interesting related observation is that long wavelength ghost instabilities can also be re-cast as tachyonic instabilities and vice versa \cite{Gumrukcuoglu:2016jbh}.} So the time scale associated with this instability in $\Lambda{}$CDM is important, as it suggests that tachyonic instabilities with similar time scales are harmless in cosmology (and can indeed be required in the context of structure formation). 
\\

\ni {\bf Stability conditions from scalar tracers}: 
As discussed above, we will now use a canonical matter scalar \eqref{Lmatter} to derive a set of stability conditions. While this will be significantly more involved for the Horndeski models considered in the following section, the GR/$\Lambda{}$CDM example discussed here will help to outline a number of important features. 
Linearly perturbing \eqref{Horndeski} and identifying ${\cal L}_m$ with \eqref{Lmatter}, we now work in spatially flat gauge and solve for the auxiliary fields $\Psi$ and $B$ 
\begin{align}
    \Psi &= \frac{\dot{\chi} \delta \chi}{2 H}, 
    &\frac{k^2}{a^2}B &= \frac{2 \dot{\chi} \delta \dot{\chi} H + \delta \chi (6 \dot{ \chi} H^2  - \dot{\chi}^3 + 2 H V')}{4 a H^2},
\end{align}
where in an abuse of notation we will denote the derivative of the potential $V$ with respect to its argument by $V'$. Substituting this back into the quadratically perturbed action, \eqref{stab_schematic} then becomes
\be \label{GRactMu}
{{\cal S}^{(2)}} = \int dx^3 dt a \big [ \delta \dot{\chi}^2 - ( \frac{k^2}{a^2} + \mu^2) \delta \chi^2 \big ].
\ee
From this expression we can already see that ghost and gradient stability conditions are trivially satisfied for $\chi$. The tachyonic stability is less trivial, however, and the corresponding effective mass $\mu^2$ in \eqref{GRactMu} is equal to
\begin{equation}
    \mu^2 = - 2 H^2 - \dot{H} + \frac{\dot{H} \dot{\chi}^2 }{2 H^2} - \frac{\dot{\chi}^4}{4 H^2} - \frac{\ddot{\chi} \dot{\chi}}{H}  + \frac{\dot{\chi} V'}{H} + V''.
 \label{GRmu2} 
\end{equation}
As already mentioned above, a tachyonic instability is present when $\mu^2 < 0$. 
%
Whether such an instability can be kept under control depends on its size and evolution time scale. \refreplybold{In order to quantify this,
we will find it useful to follow \cite{Frusciante:2019puu} and define the following dimensionless parameter}
\begin{align}
\gamma \equiv \frac{\mu^2}{H^2}.
\label{gamma}
\end{align}

A tachyonic instability is present when $\gamma < 0$ and its dimensionless amplitude measures the relative strength of such an instability with respect to the Hubble time scale.\footnote{Note that the Hubble scale is of course the typical time scale associated with cosmological evolution at large and, in particular and as discussed above, with the Jeans' instability.} Armed with this notation, we can now take \eqref{GRmu2} and explicitly compute the corresponding $\gamma$. 
Here we emphasise that \eqref{GRmu2} was computed with scalar matter proxy \eqref{Lmatter}, which one can now translate into a condition solely expressed in terms of an analogous pressure, density, their derivatives etc. Assuming that this form of the stability condition faithfully captures (part of) the relevant stability conditions for a realistic cosmological fluid, we find
\footnote{The scalar field $\chi$ and its potential $V$ are connected to the matter density and pressure in their usual manner:
\begin{align} 
\nn \rho_m &= \frac{1}{2} \dot{\chi}^2 + V[\chi], \quad   &p_m = \frac{1}{2} \dot{\chi}^2 - V[\chi].
\end{align}
In the derivation of \eqref{JM} we have assumed that $\dot\chi \neq 0$, so also $\dot H \neq 0$. An equivalent expression is
\begin{align}
\nn \gamma = &- 2 + 5 \frac{\dot{H}}{H^2} - 2 \frac{\dot{H}^2}{H^4} + 2 \frac{\ddot{H}}{H^3}  + \frac{V''(\chi)}{H^2}.
    \label{JM2}
\end{align}
This form is more useful e.g. for investigating the de Sitter limit.}
\begin{equation}
\begin{aligned}
\gamma = &- 2 + 2 \frac{\dot{H}}{H^2} - 2 \frac{\dot{H}^2}{H^4} + 2 \frac{\ddot{H}}{H^3} - \frac{2 \dddot{H}\dot{H} -  \ddot{H}^2+ 6 \ddot{H}\dot{H}H}{4 \dot{H}^2 H^2}.
\end{aligned}{}
    \label{JM}
\end{equation}
where we have used the background equations \eqref{background} in the process.
The requirement that $\gamma \geq 0$ would therefore amount to a theoretical prior demanding the absence of tachyonic instabilities, whereas requiring $\gamma \geq -|\gamma_{\rm cut}|$ is akin to excluding all instabilities larger than a given cutoff size.

To develop further intuition for the value of $\gamma$ and the corresponding `size' of any would-be tachyonic instabilities, we can evaluate the expression~(\ref{JM}) for the expansion history of a realistic cosmological model (modelled with matter and radiation fluids rather than any scalar matter proxies). Doing so in the limits of matter and radiation domination, we find
\begin{align}
\gamma_{\rm (mat)} &= -1/2,
&\gamma_{\rm (rad)} &= 0.
    \label{JMlimits}
\end{align}
To arrive at these expressions, we have used that, from~(\ref{background}), $\dot{H}/H^2$ takes the values of $-3/2$ and $-2$ during a matter and radiation domination phase, respectively. The limiting expressions in \eqref{JMlimits} agree well with the full $\gamma$ evolution over time, which we plot in fig.~\ref{JeansMass}.\footnote{Note that the late-time de Sitter limit is more subtle. The limiting value for $\gamma$ in the de Sitter limit does not smoothly connect to dark energy dominated cases with even a minimal contribution from matter. So, while evaluating $\gamma$ in this limit yields $\gamma_{\rm (dS)} = -2$, there is no tachyonic instability present in the dark energy dominated (but not exactly de Sitter) era in fig. \ref{JeansMass}.}
So a key observation is that tachyonic instabilities as measured by $\gamma$ are at most of order unity throughout the evolution. Analogous reasoning has previously led to the conjecture (see discussions in \cite{DeFelice:2016ucp,Lagos:2017hdr,Frusciante:2018vht}) that instabilities with $\gamma \gtrsim -{\cal O}(1)$ are harmless, but that theories with significantly stronger tachyonic instabilities are unviable, i.e.\footnote{
For comparison, note that we start from a slightly differently defined action than e.g. the one used in~\cite{Bellini:2014fua}: $\mathcal{S}^{(2)} = \int d^3 k dt a^3 \big [\pm (\dot \pi)^2 - (\tfrac{c_s^2 k^2}{a^2} + \mu^2) \pi^2 \big ]$. By redefining the field as $\pi \rightarrow 1/ a \pi$, it is possible to connect these two conventions. This leads to an overall shift in $\tilde{\gamma} = \gamma + 2 + \frac{\dot{H}}{H^2}$. 
} 
\begin{align} \label{conj}
{\rm \bf Conjecture}: \quad\quad \gamma \ll -1 \quad \Rightarrow \quad \text{theory unviable}.
\end{align}
This is based on the intuition that, if any such instability starts forming too quickly, this would lead to an uncontrolled growth of perturbations that ultimately comes into conflict with observational constraints. In this paper we argue that this does not need to be the case and that even `strong' tachyonic instabilities do not generically spoil the validity of the associated theories. \refreplyred{Note that a similar analysis was carried out in \cite{Frusciante:2019puu}, where tachyonic instabilities were investigated in a similar vein in the context of a generalised cubic covariant Galileon model \cite{DeFelice:2011bh}. We are therefore probing a different subset of Horndeski theories (given our choice of background and $\alpha_i$ parametrization), complementing the analysis of \cite{Frusciante:2019puu}.}
 \\

\ni {\bf Matter modelling and different scalar tracers}:
In the above we have derived stability conditions using a canonical scalar field as a matter proxy. By assuming that these conditions remain valid for a cosmological model (with many more matter degrees of freedom) we can then use and evaluate these conditions for a fully-fledged cosmological model involving matter and radiation fluids. In \cite{Lagos:2017hdr} this mapping of stability conditions was conjectured, but here we would like to more explicitly ask how robust this procedure is and what happens, if a different matter proxy is used to derive stability conditions instead. 

The example we will now consider is that of a $P(X)$ scalar field as an alternative proxy for matter. While a single canonical scalar is known to not be able to capture the behaviour of a general cosmological fluid\footnote{If we were to map a cosmology driven by a canonical matter scalar to the cosmological fluid picture described above, we would for instance encounter singularities for $\Delta$ and $c_s^2$ in the matter domination limit.} \cite{Arroja_2010,Christopherson_2009,Faraoni_2012}, 
a $P(X)$ scalar can do so more accurately\footnote{Note that there are, however, still residual singularities in the mapping between a $P(X)$ scalar and a general cosmological fluid -- see~\cite{Boubekeur:2008kn,DeFelice:2015moy} and references therein. For an alternative approach based on the Sorkin-Schutz action, that circumvents some of these issues, see 
\cite{Schutz:1977df,Brown:1992kc,DeFelice:2015moy,DeFelice:2016ucp}.}, mimicking a barotropic perfect fluid under the assumption that the fluid flow is irrotational \cite{DIEZ_TEJEDOR_2005}. To establish how robust the stability priors derived above are, it is therefore instructive to compare priors derived with a canonical matter scalar vs. those derived using a $P(X)$ scalar.
Instead of the matter Lagrangian \eqref{Lmatter} we now have
\begin{align}
{\mathcal L}_m = P(X),
\label{LPofX}
\end{align}
where $X =-\frac{1}{2} \partial_{\mu} \chi \partial^{\mu} \chi $ is the usual kinetic term for $\chi$. The (matter) stress energy tensor for this theory is
\begin{align}
     T_{\mu \nu} 
    &= \partial_{\mu} \chi \partial_{\nu} \chi P_X + g_{\mu \nu}P,
    \label{TPX}
\end{align}
where derivatives of $P$ with respect to $X$ are denoted by subscripts, e.g. $P_X \equiv \tfrac{d}{dX}P(X)$. If $\partial_{\mu} \chi$ is timelike, then the scalar field gradient acts as a natural four-velocity and we can define
\begin{equation}
    u_{\mu} \equiv  \frac{\partial_{\mu} \chi}{\sqrt{2 X}}.
\end{equation}
Using this definition and \eqref{TPX} we recover the stress energy tensor of a perfect fluid
\begin{align}
T_{\mu \nu} &= (\rho + P) u_{\mu} u_{\nu} + g_{\mu \nu} P,    &\rho &= 2 X P_X - P.
\end{align} 
where $\rho$ and $P$ denote the density and pressure of the perfect fluid.
Notice that we are somewhat abusing notation here: Since the background value of $P(X)$ corresponds to the background pressure, so we denote both with the same capital $P$ -- any $P$ appearing below will always be evaluated at the background level, so this distinction will be immaterial. Using \eqref{LPofX} we then obtain the following ghost and gradient stability conditions
\begin{align}
{\cal D} &= 2 X (P_X + 2X P_{XX}) > 0, &{\cal D}c_s^2 &= 2 X P_X \geq 0.
\end{align}
Evaluating these expressions using the background equations of motion, we find 
\begin{align}
 \mathcal{D} &= \frac{6 \dot{H}^2 H}{\ddot{H}+3 \dot{H} H}, &{\mathcal D}c_s^2 &= -2\dot H.
\end{align}
Finally, we can evaluate the expression for $\gamma$ following from \eqref{LPofX}, to be compared with \eqref{JM} (which we recall was derived using a canonical scalar as a matter proxy). Doing so we obtain
\begin{align}
 \gamma &= \frac{1}{4 H^4 (\ddot{H} + 3 \dot{H} H)^2} (6 \ddot{H}^3 H + \ddot{H}^2 (-3 \dot{H}^2 + 62 \dot{H} H^2 - 53 H^4 ) \nn \\
 &-3H^2(\dddot{H}^2 + 8 \dddot{H}\dot{H}^2 + 24 \dot{H}^4 -2 \ddddot{H} \dot{H} H - 24 \dot{H}^3 H^2 + 24 \dot{H}^2 H^4) \nn \\
 &- 2 \ddot{H} H (12 \dot{H}^3 - \ddddot{H} H - 51 \dot{H}^2 H^2 + 51 \dot{H} H^4 + \dddot{H} (\dot{H}+ 9 H^2))).
 \label{JMcomp}
\end{align}

Before comparing \eqref{JM} and \eqref{JMcomp} it is interesting to note that we can establish a mapping between the fluid and scalar variables. For example, for the comoving density contrast $\Delta$ in \eqref{DeltaEqn} we have the following mapping
\be \label{DeltaMapping}
\Delta \rightarrow \frac{2 \dot{H}(\dot{H} \delta \dot{\chi} - \dot{H} \dot{\chi} \Psi - \delta \chi (\ddot{H} + 3 \dot{H} H)))}{\dot{\chi}H(\ddot{H} + 3 \dot{H} H)},
\ee
so we can relate variables in both formulations and in particular recover the standard Jeans' instability using the above mapping. The equations of motion for $\chi$ and $\Delta$, and hence the associated stability conditions, are therefore related by the above field redefinition/mapping. Note that the mapping can be obtained by noticing that $\Delta = \delta - (3 H v_m (\rho_m + p_m))/\rho_m$, as before. The mappings for the density and pressure are given above, while for $\delta_m$ and $v_m$ we have
\begin{align}
    \delta_m &\rightarrow \frac{2 \dot{H}^2 (\delta \dot{\chi} - \dot{\chi} \Psi)}{\dot{\chi}H(\ddot{H} + 3 \dot{H} H)}, 
    &v_m &\rightarrow - \frac{\delta \chi }{\dot{\chi}}.
\end{align}
This also shows that the above mapping, while useful formally, has its limitations physically and should be used with care. It diverges in the limit of matter domination ($\ddot{H} + 3 \dot{H} H$ and hence the denominator tends to zero then), which is one of the residual singularities discussed above one encounters when insisting on a $P(X)$ model as a {\it bona fide} matter model. We therefore re-emphasise that the different matter proxies discussed here are proxies used in order to derive (some of) the relevant stability conditions, not to fully mimic the dynamical behaviour of all matter fields. 

Returning to the stability conditions, there are now three quantities we would like to compare: 1) $\gamma$ from \eqref{JM}, i.e. a tachyonic stability condition derived with \eqref{Lmatter} as a matter proxy. We will call this quantity $\gamma_{\rm can}$. 2) $\gamma$ from \eqref{JMcomp}, i.e. a tachyonic stability condition derived with \eqref{LPofX} as a matter proxy. We will call this $\gamma_{P(X)}$. 3) The coefficient of $\Delta$ in \eqref{DeltaEqn} which quantifies the strength of the standard Jeans' instability and which we will call $\gamma_\Delta$ in an abuse of notation. We plot the evolution of these three terms with redshift in fig.~\ref{JeansMass}. There are then two key observations: First, the stability conditions derived with different matter proxies are not identical -- their relative strength and presence can vary depending on the matter proxy obtained. In fact we see that the instability condition derived using a canonical scalar matter proxy is always weaker than that using a more involved $P(X)$ matter proxy (which, as discussed above, is known to mimic an overall cosmological fluid more accurately).
This corroborates our assumption above, namely that stability conditions derived using \eqref{Lmatter} are conservative tracers of the complete set of instability conditions (which can be enhanced by more accurately modelling the overall cosmological matter fluid and/or explicitly modelling additional individual components). In what follows we will therefore continue to use tachyonic instability priors derived using \eqref{Lmatter} with the assumption that they will continue to work as conservative tracers in this sense. The second important observation is that, while the presence of a standard Jeans' instability (formulated in terms of $\Delta$) and of tachyonic instabilites associated with a matter proxy (here formulated in terms of the scalar $\chi$) are indeed related by a mapping such as \eqref{DeltaMapping}, the resulting conditions themselves are not identical. The presence of one such instability at a given time therefore does not necessarily imply the presence of the other -- as shown in figure \ref{JeansMass}, the mapping and relationship is more involved.
\refreplybold{At this point also note that one can phrase the above analysis, which was carried out in a consistenly gauge-fixed manner, in a gauge-invariant language, defining a linear density perturbation $\delta_{\chi}$ as in \cite{PhysRevD.96.024060}, which then allows to study tachyonic instabilities along the lines considered here in a gauge-invariant way. Note that  in the matter modelling context of this section this gauge invariant quantity $\delta_\chi$ reduces to the fractional overdensity $\delta_m = \delta \rho_m/ \rho_m$ (as we used it for the derivation of $\gamma_\Delta$) in the Newtonian gauge.}

The above observation serves to illustrate another key point: The conditions ensuring the absence of ghost and gradient instabilities we discussed above are clear-cut, i.e. there is no convention- or formulation-dependence in the way we compute these conditions.
\refreplybold{However, the definition of an effective mass $\mu$ as in \eqref{action1SF} depends on the normalisation, i.e. different ways of normalising the relevant scalar degree of freedom will yield different effective masses and hence tachyonic stability conditions \cite{PhysRevD.96.024060}. }To gain some intuition for this, note that switching from canonically normalising in conformal time (as we do here) to doing so in physical time can introduce an ${\cal O}(1)$ shift in the corresponding $\gamma$ parameter (see footnote 7). 
More generally speaking, performing field re-definitions will generically alter tachyonic stability conditions, as illustrated by the above $\Delta$ vs. $\delta\chi$ example. So indeed tachyonic stability conditions are not as clear-cut as their ghost and gradient analogues. 
\refreplybold{Having said this, if a meaningful theoretical prior can be identified using one self-consistent set of conventions, then this can easily be mapped into a corresponding condition when expressed in terms of another field via the corresponding field re-definition. So choosing a specific convention here does not affect the generality of our results, but the precise numerical values and expressions for effective masses and stability conditions given throughout this paper ought to be understood within the context of our conventions as detailed above.}

\section{Tachyonic stability criteria} \label{sec-tach}

Having considered the (benign) nature of tachyonic instabilities in GR above, we are now in a position to put the conjecture \eqref{conj} to the test and consider it for general dark energy candidates of the form \eqref{Horndeski}. 
Importantly, while in GR there was only one propagating scalar mode associated with the matter degree of freedom $\chi$, in the dark energy models considered here there is now an additional propagating scalar mode. Since this second mode has an associated effective mass term as well, in analogy to (\ref{gamma}) we will now keep track of the following two parameters and associated tachyonic instabilities:
\begin{align}
    \gamma_m &\equiv \frac{\mu_m^2}{H^2}, 
    &\gamma_s &\equiv \frac{\mu_s^2}{H^2}.
\end{align} 
Here $\gamma_m$ is associated to the matter \dof that also propagates in GR, while $\gamma_s$ is linked to the new dark energy \dof. This second \dof is also the one linked to the scalar ghost \eqref{ghost_cond} and gradient instability conditions \eqref{sound_speed}, while the ghost and gradient stability conditions for matter \dof are trivially satisfied \cite{Lagos:2017hdr}. Having said this, note that labelling these {\it dof}s and conditions as `matter' and (dark energy) 'scalar' should not be taken too literally, since the original scalar/matter perturbations get mixed in the process of identifying the propagating {\it dof}s and deriving their associated stability criteria. \refreplyred{For the technical and explicit derivation of the $\mu_i$ (and hence $\gamma_i$), we refer the reader to \cite{DeFelice:2016ucp,PhysRevD.96.024060}, where the mass eigenvalues have been derived for the first time.}
\\

\ni{\bf Setup}:
Ultimately we are interested in whether the conjecture \eqref{conj} is true and can therefore be used as a prior in deriving constraints on the underlying model parameters. Using \eqref{conj} as a prior requires a more quantitative definition of a cutoff for $\gamma$, so we will proceed by computing cosmological parameter constraints with the following prior
\be
\begin{aligned}
\text{\bf Prior}: \quad\quad \gamma > -|\gamma_{\rm cut}|,
\end{aligned}
\label{gammaPrior}
\ee
where we will investigate different values of $\gamma_{\rm cut}$.  If for some such value $\bar\gamma_{\rm cut}$, constraints derived with this prior exclude a significant part of parameter space that {\it does not} yield good fits to observations (so the prior does indeed have a useful effect), while simultaneously not excluding any regions of parameter space that {\it do} yield good fits (i.e. we do not want the prior to be overzealous and exclude perfectly valid regions of parameter space), then \eqref{gammaPrior} with $\gamma_{\rm cut} = \bar\gamma_{\rm cut}$ is indeed a useful physical prior to implement. Regarding this second point (avoiding overzealous cuts), we emphasise that the presence of a tachyonic instability, no matter how strong, is only a problem if its presence is correlated to badly behaved perturbations and hence bad fits to the data. In other words, in our present context \eqref{gammaPrior} should not be seen as a prior motivated by some other underlying and more fundamental reason (that goes beyond observables). 

Also note that we will apply the above prior \eqref{gammaPrior} during dark energy and matter dominated epochs, so in effect for redshifts $z \lesssim 3000$. 
While tachyonic instabilities at earlier times are also of interest, e.g. related to setting well-defined initial conditions for dark energy scalars in Einstein-Boltzmann solvers~\cite{Bellini:2019syt}, this conservative choice is partially motivated by the known potential presence of observationally inconsequential instabilities in the radiation-dominated epoch, where the dark energy \dof is highly subdominant and where such instabilities can be a consequence of limitations of the simple parametrisation used \eqref{param1} rather than of underlying physical issues \cite{Noller:2020afd,radstab}. We leave a more detailed investigation of early Universe tachyonic instabilities, complementary to the late time exploration carried out here, for future work.
\\

\ni {\bf Tachyonic instabilities and cosmological constraints}:
Having set up the problem as described above, we now compute the cosmological evolution and resulting constraints on dark energy models as specified by \eqref{Horndeski}, \eqref{alphas} and \eqref{param1}. In figure \ref{fig-gammas} we show the evolution of $\gamma$ for several example models. The instability is always present in the matter sector, as expected and indeed required -- see our discussion regarding the Jeans' instability in the previous section. Deep in matter domination $\gamma_m$ is effectively identical for GR and the dark energy cosmologies, since the effect of the extra degree of freedom is strongly suppressed there, but upon approaching the dark energy dominated regime, also $\gamma_m$ is modified by the presence of the dark energy scalar and relatively strong $\gamma < -1$ instabilities can be reached easily. For the scalar dark energy sector (absent in pure GR), figure \ref{fig-gammas} shows that tachyonic instabilities are absent altogether for some parameter choices, while relatively strong instabilities can be triggered in the dark energy dominated epoch.\footnote{Note that figure \ref{fig-gammas} also serves as a consistency check, as we compute the $\gamma_i$ both using a hi\_class and an independent analytic implementation.} 

In figure \ref{fig-tachyRegion} we then show the regions in the $c_i$ parameter space that would be excluded by various $\gcut$ priors. For each point in this parameter space we compute the evolution of the $\gamma_i$ in the way illustrated in figure \ref{fig-gammas} and, if $\gamma$ violates the prior \eqref{gammaPrior} at any point in the evolution, we mark the corresponding region in parameter space as excluded. Note that we consider cases when the prior is applied to just the dark energy scalar sector (upper row) in figure \ref{fig-tachyConstraints} as well as for the case when it is applied to both dark energy scalar and matter sectors, i.e. to both $\gamma_s$ and $\gamma_m$ (lower row). Clearly the prior excludes less and less parameter space as $\gcut$ grows, ceasing to have any effect for the observationally relevant region for $\gcut \gtrsim 100$. Also note that the presence of a Jeans-like instability manifests itself by the fact that the whole region shown for $\gcut \leq 1$ is excluded when also applying the prior to $\gamma_m$. That this does not only happen for $\gcut = 0$, but also for $\gcut = 1$ suggests that this Jeans-like instability in the matter sector is generically stronger in the presence of a dark energy scalar as modelled here than it is in pure $\Lambda{}$CDM. This is not altogether surprising, since we have worked with a non-trivial kineticity (setting $c_k= 0.1$) throughout most of this paper (for a comment on the $\alpha_k$ dependence of the tachyonic instability, see appendix~\ref{appendix_aK}). This generically affects tachyonic instabilities for both scalar and matter modes, so even when considering the limit $\{\alpha_M \to 0, \alpha_B \to 0\}$, there is still a non-trivial dependence on the dark energy scalar and one should therefore not expect to recover pure $\Lambda{}$CDM behaviour here.

In figure \ref{fig-tachyConstraints} we then explicitly compute cosmological parameter constraints for a variety of $\gcut$ priors. The resulting contours neatly match up with the parameter space regions shown in figure \ref{fig-tachyRegion}. We find that the full observationally acceptable region is recovered only for $\gcut \gtrsim 100$.
For smaller choices of $\gcut$ this new prior always results in overzealous cuts eliminating perfectly viable regions of parameter space. Indeed this interestingly happens particularly for cosmologies with very small $c_i$, i.e. very close to (but not identical to) GR predictions -- so it is important not to erroneously exclude such regions due to a hard tachyonic stability prior.
In particular note that this also implies that tachyonic instabilities which are significantly `stronger' (as measured by $\gamma$) than the Jeans instability in GR do not spoil the observational validity of the theory and in fact comfortably sit within the $1\sigma$ region most favoured by observational constraints.
Returning to $\gcut \gtrsim 100$, while the full observationally viable parameter space is recovered for such priors, a comparison with figure \ref{fig-tachyRegion} shows that the prior then in fact does not exclude any relevant regions of parameter space at all, so it simply is irrelevant in this case. 
Compare this with the relevant gradient stability prior as shown in figure \ref{gradientcut}. There parameter constraints obtained with and without applying this prior identified the same resulting valid region, but gradient stability priors helped increase the efficiency of the sampling by excluding regions {\it a priori} that are in close proximity to the observationally viable parts of parameter space. In the absence of a gradient stability prior, these regions would have been extensively sampled by an MCMC exploration, but would then have been excluded by the data {\it a posteriori}. Placing a gradient stability prior was therefore useful without biasing the final result, whereas for the tachyonic stability priors above we either see strong biasing or no useful impact at all.
This strongly cautions against applying tachyonic stability priors, showing that the conjecture \eqref{conj} fails in the setups considered throughout this paper.
\\

\section{Conclusions} \label{sec-conclusions}

In this paper we investigated tachyonic instabilities in dark energy theories and to what extent priors related to (requiring the absence of) these instabilities can play a useful and informative role in the extraction of cosmological parameter constraints. As discussed in section \ref{sec-grad}, well-established priors related to ghost and gradient instabilities significantly increase the efficiency of constraint extraction, without significantly biasing the eventual result (by which we here mean: without mistakenly ruling out viable regions of parameter space). Tachyonic instabilities are significantly more subtle in that their mere presence is clearly not the sign of an underlying sickness in the theory -- the Jeans' instability in GR is a primary example, as discussed in section \ref{sec-Jeans}. 
Motivated by this, it had been conjectured (see e.g. discussions in \cite{DeFelice:2016ucp,Lagos:2017hdr,Frusciante:2018vht}), that the presence of sufficiently strong tachyonic instabilities (in particular, stronger than the Jeans' instability) can be used as a diagnostic to detect unviable theories. 
Here we therefore investigated the evolution of the effective mass of cosmological perturbations and of the closely linked tachyonic instabilities for general Horndeski gravity theories in detail, attempting to identify a well-motivated cutoff that demarcates strong from acceptable tachyonic instabilities. Having computed cosmological constraints for a range of such candidate cutoffs, we conclude that the conjecture ultimately fails in the present context. The cutoff generically is either overzealous and excludes perfectly viable regions of parameter space, or it is so weak that it has next to no effect on the extraction of cosmological parameter constraints. This suggests that, while there may be specific examples with fine-tuned matching priors that can side-step these worries, priors based on excluding cosmologies with sufficiently strong tachyonic instabilities can be safely ignored for general dark energy models.  \refreplyred{Note that our analysis here complements that of \cite{Frusciante:2019puu}, where similar conclusions were found in the context of a generalised cubic covariant Galileon model \cite{DeFelice:2011bh}.}

We close by emphasising that several complementary theoretical priors beyond those considered here exist, which it will be interesting to further explore and include in future investigations of the interplay between theoretical priors and observational constraints. Here we have focused on the relatively well-investigated subset of classical (ghost, gradient and tachyonic) stability criteria for scalar modes propagating on cosmological backgrounds. Interesting complementary classical stability priors come e.g. from
considering the propagation of dark energy perturbations on backgrounds sourced by binary mergers, yielding a constraint on the size and presence of gravitational-wave induced dark energy instabilities \cite{Creminelli:2019kjy} (also see the closely related \cite{Creminelli:2019nok,Creminelli:2018xsv}). This in turn significantly further tightens cosmological parameter constraints on dark energy \cite{Noller:2020afd}. In this context also note complementary constraints tightly linking local solar system constraints to cosmology \cite{Babichev:2011iz,Burrage:2020jkj,Noller:2020lav}. 
Similarly, requiring radiative (rather than just classical) stability can impose additional constraints on dark energy theories (for more detailed discussions see \cite{radstab,Heisenberg:2020cyi} and references therein), with the `weakly broken Galileon' \cite{Pirtskhalava:2015nla} (i.e. shift symmetric Horndeski theories) a well-known example of cosmologically motivated scalar-tensor theories with parametrically suppressed radiative corrections. 
All of the above are constraints directly diagnosable at the level of the low energy effective theories describing dark energy at large scales (such as Horndeski gravity). Demanding that such theories have a sensible UV completion can constrain them yet further -- see \cite{Melville:2019wyy,Kennedy:2020ehn,deRham:2021fpu} and references therein for a discussion of how the resulting bounds can affect the cosmological parameter constraints discussed here. 
All these different theoretical priors constrain linear cosmology (and gravitational physics at large) in a variety of powerful and orthogonal ways.
Exploring additional candidate (theoretically or observationally motivated) priors in order to better understand dark energy, as we have done here in the context of tachyonic instabilities, will therefore play an essential role in obtaining ever tighter constraints going forward.

\acknowledgments
We especially thank Emilio Bellini for several very helpful discussions and shared insights.
We also thank David Bacon, Rob Crittenden, Emir G\"umr\"uk\c{c}\"uoglu, Kazuya Koyama and Alexandre Refregier for useful discussions and comments on a draft.
RG is supported by the PhD program of the University of
Portsmouth.
JN is supported by an STFC Ernest Rutherford Fellowship, grant reference ST/S004572/1, and also acknowledges support from King's College Cambridge, Dr. Max R\"ossler, the Walter Haefner Foundation and the ETH Zurich Foundation.  
In deriving the results of this paper, we used: CLASS \cite{Blas:2011rf},  corner \cite{corner}, hi\_class \cite{Zumalacarregui:2016pph,Bellini:2019syt}, MontePyton \cite{Audren:2012wb,Brinckmann:2018cvx} and xAct \cite{xAct}.
\\

\appendix

\section{The $\alpha_K$--dependence of tachyonic instabilities} \label{appendix_aK}
In the main text we have not explicitly discussed any dependence of the tachyonic stability prior \eqref{gammaPrior} on $\aK$. We have done so (and fixed $c_k$ to a fiducial value in the process) for good reason, since it is by now well-understood, that cosmological parameter constraints at most very weakly depend on $\aK$\cite{BelliniParam}.
This is closely related to the fact that $\aK$ drops out of the dynamics controlling linearised cosmological perturbations at leading order in the quasi-static regime \cite{BelliniParam, Alonso:2016suf},
which applies on all but the very largest scales (whose constraining power is weakened by cosmic variance).
The effective mass term $\mu^2$ that we have focused on here is of course sub-dominant in the quasi-static approximation itself. So phrased in this way the motivation of the conjecture \eqref{conj} was that, despite its subdominant contribution to observable scales, instabilities linked to the effective mass term would nevertheless correlate with relevant regions in parameter space that poorly fit the data. However, we have seen that this is not the case above and explicitly considering the dependence of tachyonic instabilities on $\aK$ will offer a different perspective on why this is the case.

\begin{figure}[t!]
\includegraphics[width=\linewidth]{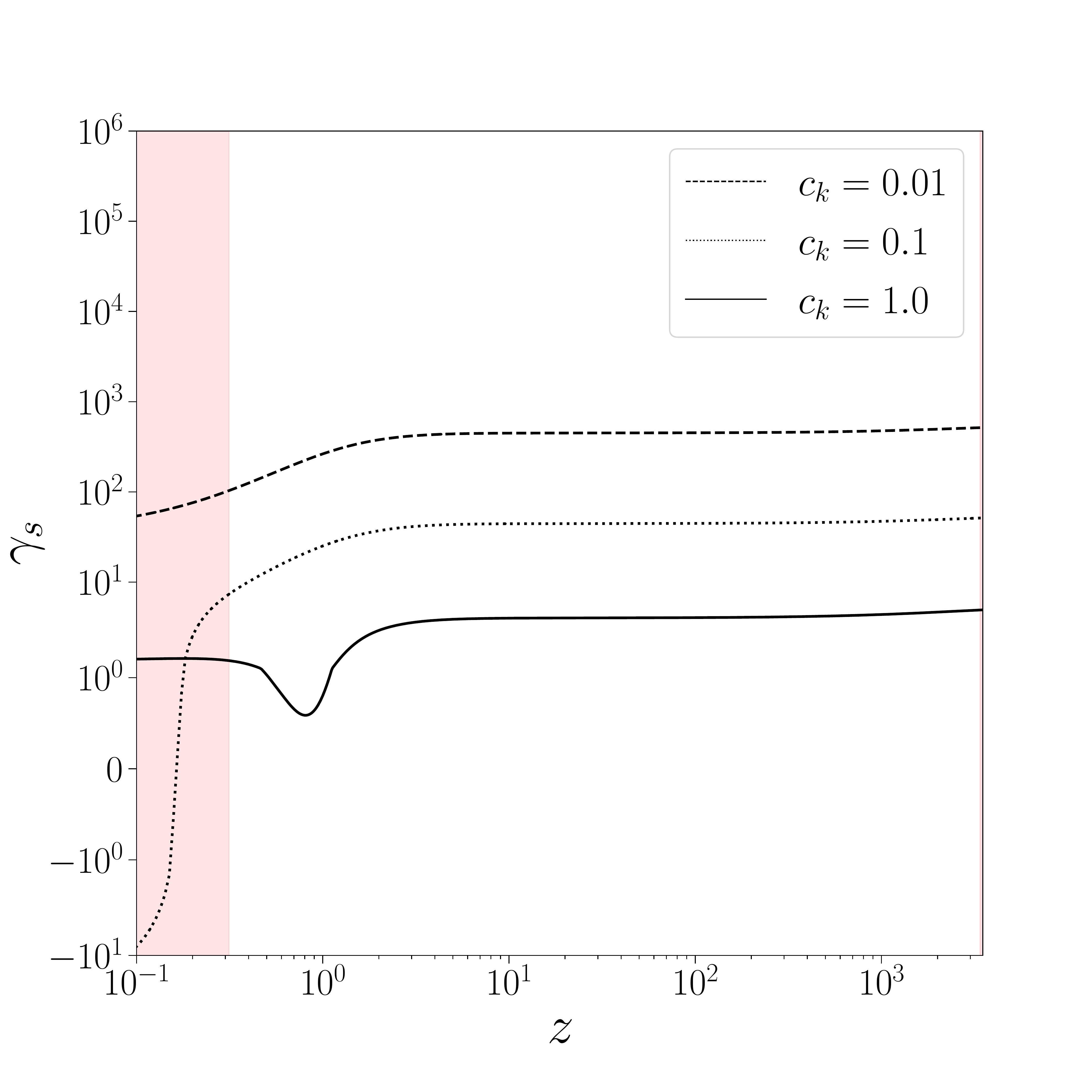}
\caption{
Evolution of the parameter $\gamma_s$ for three different fiducial values of $c_k$ and fixed $c_b = 0, c_m = 1.0$.
This clearly shows that the (fiducial) choice of $\alpha_K$ can amplify/reduce the overall size and evolution of tachyonic instabilities. Especially notice that the instabilities do not scale linearly with $c_k$.
\label{fig-aKevolutions}}
\end{figure}

\begin{figure*}[t!]
\includegraphics[width=\linewidth]{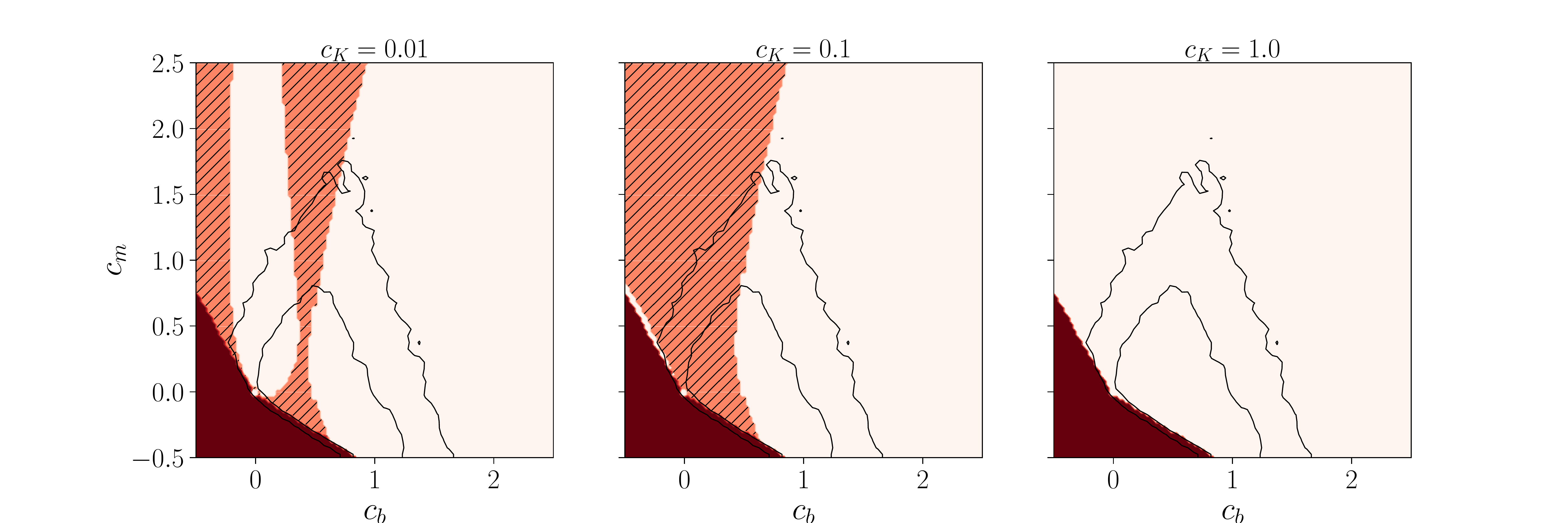}
\caption{
Here we show the analogue of figure \ref{fig-tachyRegion}, where we only apply $\gamma_{\rm cut} = 1$ to $\gamma_s$, i.e. to the dark energy scalar sector.
As before the shading denotes the presence of different instabilities: gradient (solid), tachyonic with $\gamma_s > 1$ (striped) or neither of the previous two cases  (no shading). Importantly different choices of $c_k$ (and hence $\alpha_K$) significantly affect the presence and size of tachyonic instabilities. 
\label{fig-aKregions}}
\end{figure*}

While $\aK$ does not affect the gradient stability condition linked to \eqref{sound_speed}, it is bounded by the no-ghost condition \eqref{ghost_cond} and, for our purposes most importantly, does affect the effective mass term $\mu^2$ (non-linearly) and hence the potential presence of tachyonic instabilities.
\refreplybold{Note that $\aK$ therefore affects the shape of the overall parameter space explored by the MCMC sampler, as pointed out by \cite{Kreisch:2017uet,Frusciante:2018jzw}. This aspect may be relevant in analysing models without $\Lambda{}$CDM limits (i.e. models different from the ones explored throughout this paper).} 
In figure \ref{fig-aKevolutions} we provide a few examples of how the evolution of $\mu^2$ for the dark energy scalar mode (and hence of $\gamma_s$) is affected by changing $\aK$. Already from these examples, we see that $\gamma_s$ non-linearly depends on $\aK$. More specifically, it is not just the size, but more importantly the presence of a tachyonic instability itself that non-linearly depends on $\aK$.
Figure \ref{fig-aKregions} then shows that this is not just an artefact of the specific examples shown before, but that $\aK$ significantly affects the size and presence of tachyonic instabilities in general. In particular note that, as observed in figure \ref{fig-aKevolutions} before, for a given point in the $\aM-\aB$ plane, the size and/or presence of any would-be tachyonic instability does not scale linearly with $\aK$. More specifically, by altering $\aK$, sizeable instabilites can also be triggered and amplified for mild and observationally viable departures from GR.  
Since we know that the value of $\aK$ does not significantly affect observational constraints, this is again in conflict with the conjecture \eqref{conj}. So, as before, we conclude that no tachyonic stability prior of the type discussed here should be applied to a cosmological constraint analysis.\\

\section{Stability constraints for other parameterisations} \label{appendix_param}
\begin{figure*}[t]
\includegraphics[width=.45\linewidth]{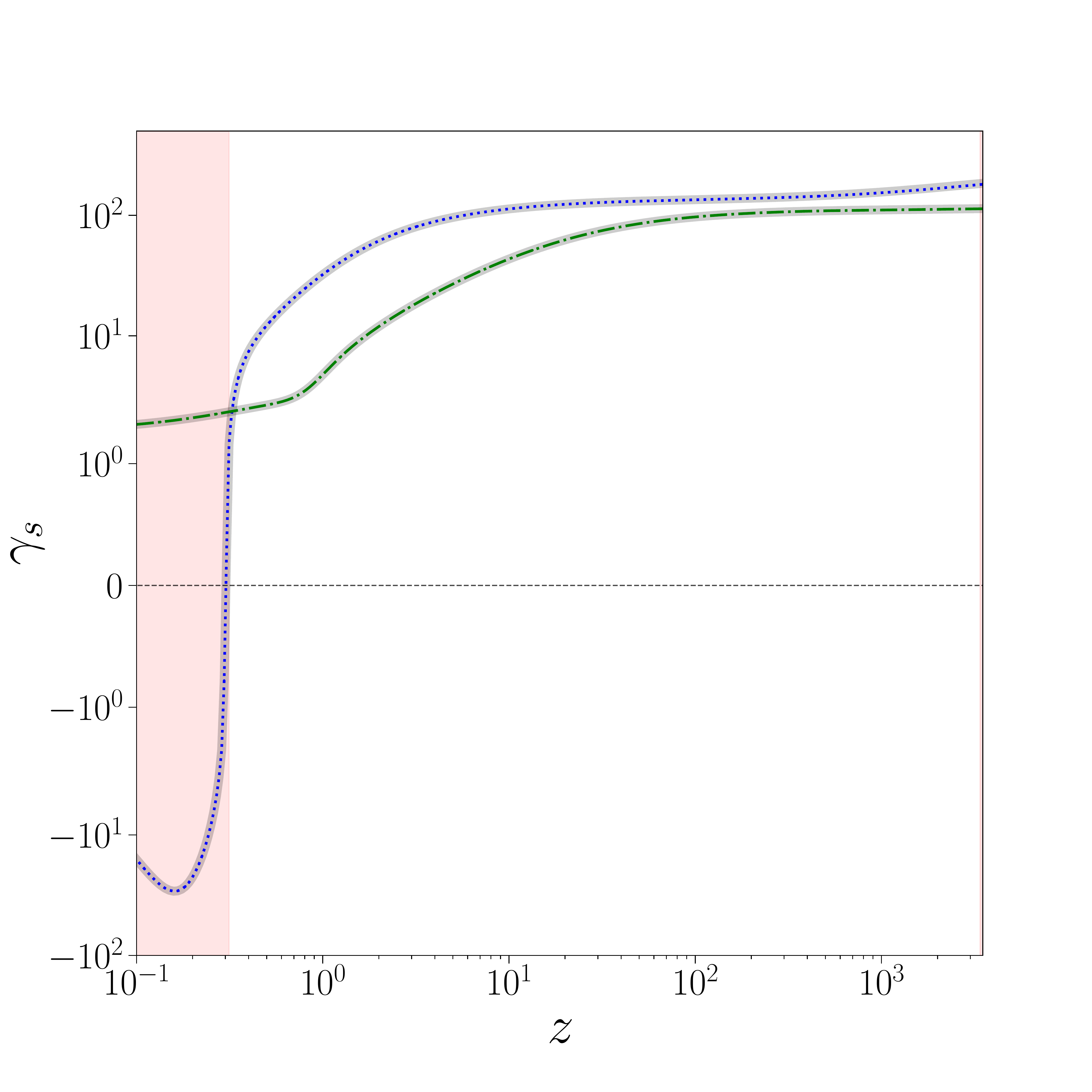}
\includegraphics[width=.45\linewidth]{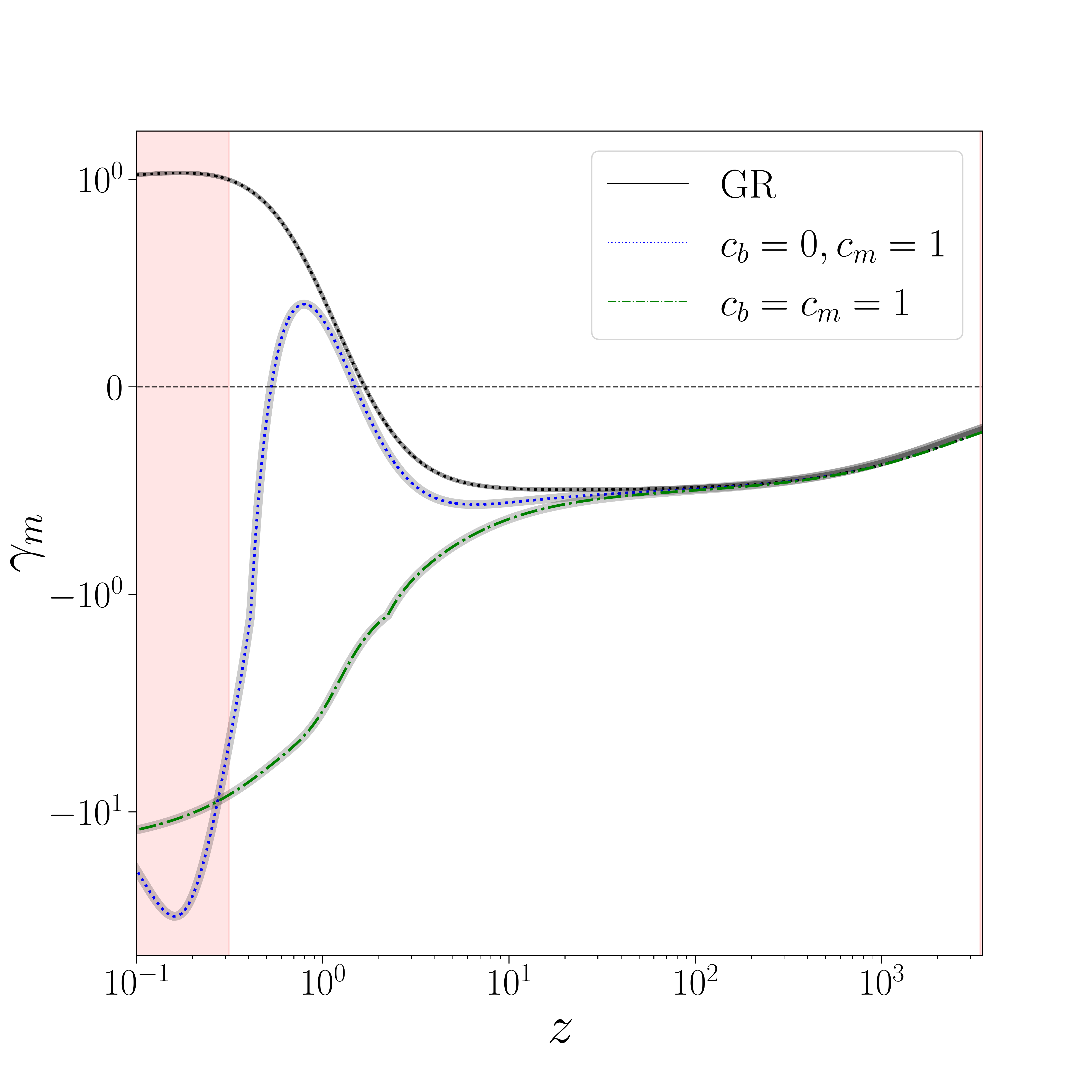}
\caption{
Evolution of the dimensionless parameters $\gamma_s$ and $\gamma_m$ over redshift, where $c_k$ is fixed to 0.1 and $c_b$ and $c_m$ are varied according to the legend. Unlike in the main text we here choose $\alpha_i = c_i \cdot a$ as an alternative parametrisations for the $\alpha$'s. 
As before, our {\it hi\_class}-based implementation (fainter,broader lines) is in excellent agreement with the independent analytical cross-checks (solid/dotted/dashed/dashed-dotted lines). The shaded region on the left corresponds to the dark energy dominated era.
Unlike for the $\alpha_i \propto \Omega$ parameterisation shown in figure \ref{fig-gammas}, here we do not compute a cosmology with $c_b= 1, c_m = 0$ (this is so far from the viable region of parameter space, no sensible initial conditions can be set for this case).
\label{vsmathematica_scale}}
\end{figure*}

\begin{figure*}[t]
\includegraphics[width=\linewidth]{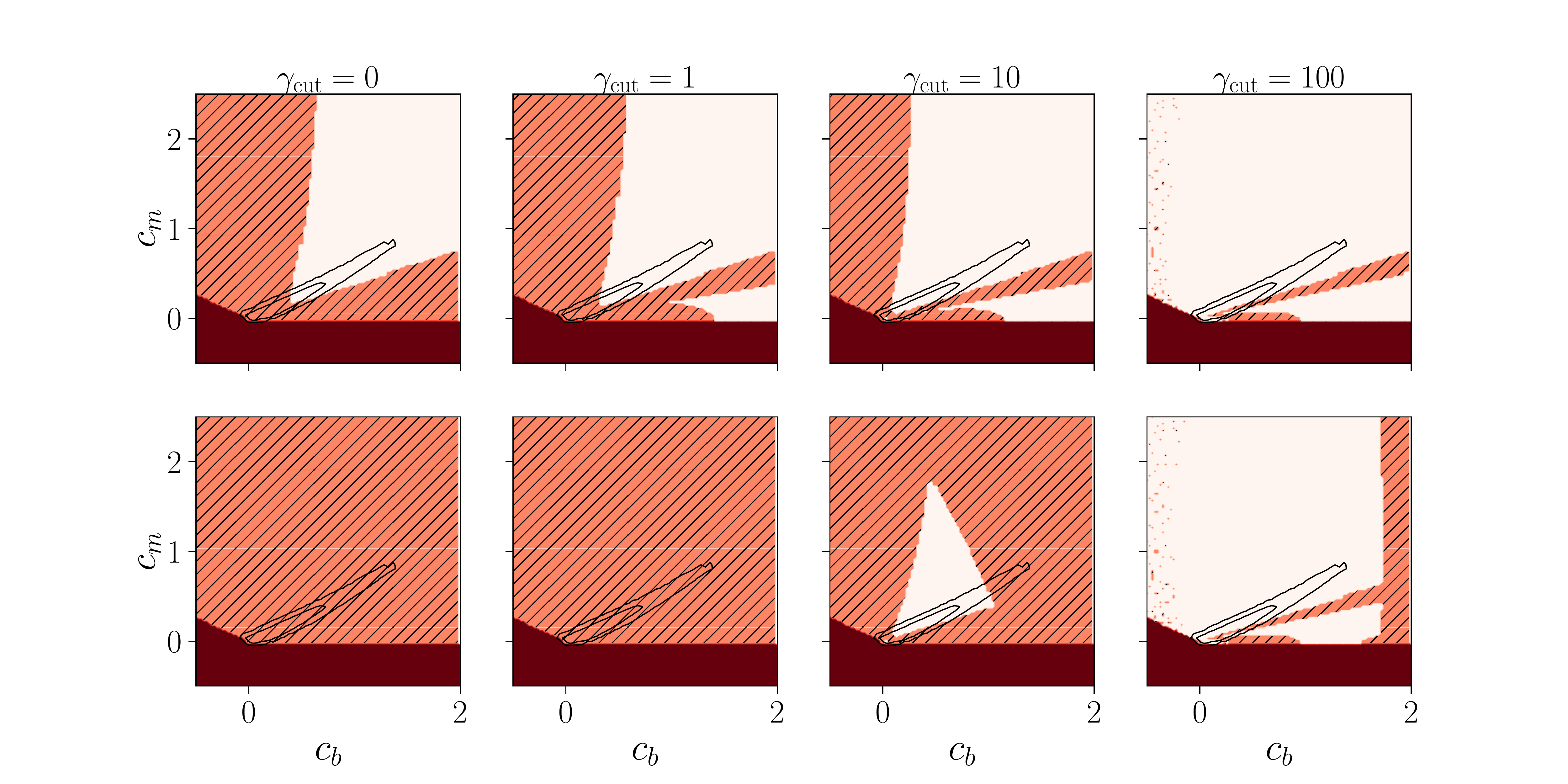}
\caption{Analogous plot to figure \ref{fig-tachyRegion},
but where we here choose $\alpha_i = c_i \cdot a$ as an alternative parametrisations for the $\alpha$'s. 
The black contours correspond to cosmological parameter constraints generated without tachyonic stability priors, with inner (outer) contours corresponding to $68 \% \; ( 95 \% )$ confidence levels -- for more detail on those observational constraints see \cite{mcmc}. When comparing with those constraints, we find that just as for the $\alpha$ parametrisation used in the main text, large observationally viable regions of parameter space are incorrectly excluded with a hard $\gamma_{\rm cut} \lesssim 10$ prior. For the parametrisation shown here this also remains true (albeit to a much smaller extent) for priors with larger $\gamma_{\rm cut}$.
\label{Meff_scale}}
\end{figure*}

Throughout this paper we have focused on stability conditions for the $\alpha_i$ in terms of a dark energy density parameterisation~(\ref{param1}). In the following we show the results of the stability constraints analysis for the parameterisation $\alpha_i = c_i a$.
In analogy to figure~\ref{fig-gammas}, figure~\ref{vsmathematica_scale} displays the comparison of the numerical evolution (fainter, broader lines) vs. the analytical calculation (solid/dashed/dotted lines) of the parameters $\gamma_s$ and $\gamma_m$ for different cosmologies. We see that the code is able to reproduce the analytical solutions in very good agreement.  The cosmology $c_b= 1, c_m = 0$ is not listed, since the code had difficulties to reproduce the analytical expressions. But in the newest {\it hi\_class} version, this cosmology is excluded by the IC tests (tachyonic instability in the radiation dominated era) anyway. In figure~\ref{Meff_scale}, the parameter space of $c_b$ and $c_m$ is plotted for different values of $\gamma_{\rm cut}$. Since both parameterisations make sure, that the dark energy perturbations become important just at late time, figures~\ref{fig-tachyRegion} and~\ref{Meff_scale} show an overall similar behaviour --
see~\cite{mcmc} for the corresponding observational constraints computed for the $\alpha_i = c_i a$ parameterisation.

\bibliographystyle{utphys}
\bibliography{tachStab}

\providecommand{\href}[2]{#2}\begingroup\raggedright\begin{thebibliography}{10}

\bibitem{Zumalacarregui:2016pph}
M.~Zumalacárregui, E.~Bellini, I.~Sawicki, J.~Lesgourgues, and P.~G. Ferreira,
  ``{hi\_class: Horndeski in the Cosmic Linear Anisotropy Solving System},''
  \href{http://dx.doi.org/10.1088/1475-7516/2017/08/019}{{\em JCAP} {\bf 1708}
  (2017) no.~08, 019},
\href{http://arxiv.org/abs/1605.06102}{{\tt arXiv:1605.06102 [astro-ph.CO]}}.

\bibitem{Hu:2013twa}
B.~Hu, M.~Raveri, N.~Frusciante, and A.~Silvestri, ``{Effective Field Theory of
  Cosmic Acceleration: an implementation in CAMB},''
  \href{http://dx.doi.org/10.1103/PhysRevD.89.103530}{{\em Phys. Rev.} {\bf
  D89} (2014) no.~10, 103530},
\href{http://arxiv.org/abs/1312.5742}{{\tt arXiv:1312.5742 [astro-ph.CO]}}.

\bibitem{mcmc}
J.~Noller and A.~Nicola, ``{Cosmological parameter constraints for Horndeski
  scalar-tensor gravity},''
  \href{http://dx.doi.org/10.1103/PhysRevD.99.103502}{{\em Phys. Rev. D} {\bf
  99} (2019) no.~10, 103502}, \href{http://arxiv.org/abs/1811.12928}{{\tt
  arXiv:1811.12928 [astro-ph.CO]}}.

\bibitem{BelliniParam}
E.~{Bellini}, A.~J. {Cuesta}, R.~{Jimenez}, and L.~{Verde},
  \href{http://dx.doi.org/10.1088/1475-7516/2016/02/053}{``{Constraints on
  deviations from {$\Lambda$}CDM within Horndeski gravity},''{\em JCAP} {\bf 2}
  (Feb., 2016)  053}, \href{http://arxiv.org/abs/1509.07816}{{\tt
  arXiv:1509.07816}}.

\bibitem{Raveri:2014cka}
M.~Raveri, B.~Hu, N.~Frusciante, and A.~Silvestri, ``{Effective Field Theory of
  Cosmic Acceleration: constraining dark energy with CMB data},''
  \href{http://dx.doi.org/10.1103/PhysRevD.90.043513}{{\em Phys. Rev.} {\bf
  D90} (2014) no.~4, 043513},
\href{http://arxiv.org/abs/1405.1022}{{\tt arXiv:1405.1022 [astro-ph.CO]}}.

\bibitem{Gleyzes:2015rua}
J.~Gleyzes, D.~Langlois, M.~Mancarella, and F.~Vernizzi, ``{Effective Theory of
  Dark Energy at Redshift Survey Scales},''
  \href{http://dx.doi.org/10.1088/1475-7516/2016/02/056}{{\em JCAP} {\bf 1602}
  (2016) no.~02, 056},
\href{http://arxiv.org/abs/1509.02191}{{\tt arXiv:1509.02191 [astro-ph.CO]}}.

\bibitem{Kreisch:2017uet}
C.~D. Kreisch and E.~Komatsu, ``{Cosmological Constraints on Horndeski Gravity
  in Light of GW170817},''
\href{http://arxiv.org/abs/1712.02710}{{\tt arXiv:1712.02710 [astro-ph.CO]}}.

\bibitem{Alonso:2016suf}
D.~Alonso, E.~Bellini, P.~G. Ferreira, and M.~Zumalacarregui, ``{Observational
  future of cosmological scalar-tensor theories},''
  \href{http://dx.doi.org/10.1103/PhysRevD.95.063502}{{\em Phys. Rev.} {\bf
  D95} (2017) no.~6, 063502},
\href{http://arxiv.org/abs/1610.09290}{{\tt arXiv:1610.09290 [astro-ph.CO]}}.

\bibitem{Arai:2017hxj}
S.~Arai and A.~Nishizawa, ``{Generalized framework for testing gravity with
  gravitational-wave propagation. II. Constraints on Horndeski theory},''
  \href{http://dx.doi.org/10.1103/PhysRevD.97.104038}{{\em Phys. Rev.} {\bf
  D97} (2018) no.~10, 104038},
\href{http://arxiv.org/abs/1711.03776}{{\tt arXiv:1711.03776 [gr-qc]}}.

\bibitem{Frusciante:2018jzw}
N.~Frusciante, S.~Peirone, S.~Casas, and N.~A. Lima, ``{The road ahead of
  Horndeski: cosmology of surviving scalar-tensor theories},''
\href{http://arxiv.org/abs/1810.10521}{{\tt arXiv:1810.10521 [astro-ph.CO]}}.

\bibitem{Reischke:2018ooh}
R.~Reischke, A.~Spurio~Mancini, B.~M. Sch\"afer, and P.~M. Merkel,
  ``{Investigating scalar\textendash{}tensor gravity with statistics of the
  cosmic large-scale structure},''
  \href{http://dx.doi.org/10.1093/mnras/sty2919}{{\em Mon. Not. Roy. Astron.
  Soc.} {\bf 482} (2019) no.~3, 3274--3287}.

\bibitem{Mancini:2018qtb}
A.~Spurio~Mancini, R.~Reischke, V.~Pettorino, B.~M. Schäfer, and
  M.~Zumalacárregui, ``{Testing (modified) gravity with 3D and tomographic
  cosmic shear},'' \href{http://dx.doi.org/10.1093/mnras/sty2092}{{\em Mon.
  Not. Roy. Astron. Soc.} {\bf 480} (2018)  3725},
\href{http://arxiv.org/abs/1801.04251}{{\tt arXiv:1801.04251 [astro-ph.CO]}}.

\bibitem{Brando:2019xbv}
G.~Brando, F.~T. Falciano, E.~V. Linder, and H.~E.~S. Velten, ``{Modified
  gravity away from a $\Lambda$CDM background},''
  \href{http://dx.doi.org/10.1088/1475-7516/2019/11/018}{{\em JCAP} {\bf 11}
  (2019)  018}, \href{http://arxiv.org/abs/1904.12903}{{\tt arXiv:1904.12903
  [astro-ph.CO]}}.

\bibitem{Arjona:2019rfn}
R.~Arjona, W.~Cardona, and S.~Nesseris, ``{Designing Horndeski and the
  effective fluid approach},''
  \href{http://dx.doi.org/10.1103/PhysRevD.100.063526}{{\em Phys. Rev. D} {\bf
  100} (2019) no.~6, 063526}, \href{http://arxiv.org/abs/1904.06294}{{\tt
  arXiv:1904.06294 [astro-ph.CO]}}.

\bibitem{Raveri:2019mxg}
M.~Raveri, ``{Reconstructing Gravity on Cosmological Scales},''
  \href{http://dx.doi.org/10.1103/PhysRevD.101.083524}{{\em Phys. Rev. D} {\bf
  101} (2020) no.~8, 083524}, \href{http://arxiv.org/abs/1902.01366}{{\tt
  arXiv:1902.01366 [astro-ph.CO]}}.

\bibitem{Perenon:2019dpc}
L.~Perenon, J.~Bel, R.~Maartens, and A.~de~la Cruz-Dombriz, ``{Optimising
  growth of structure constraints on modified gravity},''
  \href{http://dx.doi.org/10.1088/1475-7516/2019/06/020}{{\em JCAP} {\bf 1906}
  (2019)  020},
\href{http://arxiv.org/abs/1901.11063}{{\tt arXiv:1901.11063 [astro-ph.CO]}}.

\bibitem{Frusciante:2019xia}
N.~Frusciante and L.~Perenon, ``{Effective field theory of dark energy: A
  review},'' \href{http://dx.doi.org/10.1016/j.physrep.2020.02.004}{{\em Phys.
  Rept.} {\bf 857} (2020)  1--63}, \href{http://arxiv.org/abs/1907.03150}{{\tt
  arXiv:1907.03150 [astro-ph.CO]}}.

\bibitem{SpurioMancini:2019rxy}
A.~Spurio~Mancini, F.~K\"ohlinger, B.~Joachimi, V.~Pettorino, B.~M. Sch\"afer,
  R.~Reischke, E.~van Uitert, S.~Brieden, M.~Archidiacono, and J.~Lesgourgues,
  ``{KiDS + GAMA: constraints on horndeski gravity from combined large-scale
  structure probes},'' \href{http://dx.doi.org/10.1093/mnras/stz2581}{{\em Mon.
  Not. Roy. Astron. Soc.} {\bf 490} (2019) no.~2, 2155--2177},
  \href{http://arxiv.org/abs/1901.03686}{{\tt arXiv:1901.03686 [astro-ph.CO]}}.

\bibitem{Bonilla:2019mbm}
A.~Bonilla, R.~D'Agostino, R.~C. Nunes, and J.~C.~N. de~Araujo, ``{Forecasts on
  the speed of gravitational waves at high $z$},''
  \href{http://dx.doi.org/10.1088/1475-7516/2020/03/015}{{\em JCAP} {\bf 03}
  (2020)  015}, \href{http://arxiv.org/abs/1910.05631}{{\tt arXiv:1910.05631
  [gr-qc]}}.

\bibitem{Baker:2020apq}
T.~Baker and I.~Harrison, ``{Constraining Scalar-Tensor Modified Gravity with
  Gravitational Waves and Large Scale Structure Surveys},''
  \href{http://dx.doi.org/10.1088/1475-7516/2021/01/068}{{\em JCAP} {\bf 01}
  (2021)  068}, \href{http://arxiv.org/abs/2007.13791}{{\tt arXiv:2007.13791
  [astro-ph.CO]}}.

\bibitem{Joudaki:2020shz}
S.~Joudaki, P.~G. Ferreira, N.~A. Lima, and H.~A. Winther, ``Testing gravity on
  cosmic scales: A case study of jordan-brans-dicke theory,''
  \href{http://dx.doi.org/10.1103/PhysRevD.105.043522}{{\em Phys. Rev. D} {\bf
  105} (2022) no.~4, 043522}.

\bibitem{Noller:2020lav}
J.~Noller, L.~Santoni, E.~Trincherini, and L.~G. Trombetta, ``{Scalar-tensor
  cosmologies without screening},''
  \href{http://dx.doi.org/10.1088/1475-7516/2021/01/045}{{\em JCAP} {\bf 01}
  (2021)  045}, \href{http://arxiv.org/abs/2008.08649}{{\tt arXiv:2008.08649
  [astro-ph.CO]}}.

\bibitem{Noller:2020afd}
J.~Noller, ``{Cosmological constraints on dark energy in light of gravitational
  wave bounds},'' \href{http://dx.doi.org/10.1103/PhysRevD.101.063524}{{\em
  Phys. Rev. D} {\bf 101} (2020) no.~6, 063524},
  \href{http://arxiv.org/abs/2001.05469}{{\tt arXiv:2001.05469 [astro-ph.CO]}}.

\bibitem{radstab}
J.~Noller and A.~Nicola, ``{Radiative stability and observational constraints
  on dark energy and modified gravity},''
  \href{http://dx.doi.org/10.1103/PhysRevD.102.104045}{{\em Phys. Rev. D} {\bf
  102} (2020) no.~10, 104045}, \href{http://arxiv.org/abs/1811.03082}{{\tt
  arXiv:1811.03082 [astro-ph.CO]}}.

\bibitem{DeFelice:2016ucp}
A.~De~Felice, N.~Frusciante, and G.~Papadomanolakis, ``{On the stability
  conditions for theories of modified gravity in the presence of matter
  fields},'' \href{http://dx.doi.org/10.1088/1475-7516/2017/03/027}{{\em JCAP}
  {\bf 1703} (2017) no.~03, 027},
\href{http://arxiv.org/abs/1609.03599}{{\tt arXiv:1609.03599 [gr-qc]}}.

\bibitem{Lagos:2017hdr}
M.~Lagos, E.~Bellini, J.~Noller, P.~G. Ferreira, and T.~Baker, ``{A general
  theory of linear cosmological perturbations: stability conditions, the
  quasistatic limit and dynamics},''
  \href{http://dx.doi.org/10.1088/1475-7516/2018/03/021}{{\em JCAP} {\bf 1803}
  (2018) no.~03, 021},
\href{http://arxiv.org/abs/1711.09893}{{\tt arXiv:1711.09893 [gr-qc]}}.

\bibitem{Frusciante:2018vht}
N.~Frusciante, G.~Papadomanolakis, S.~Peirone, and A.~Silvestri, ``{The role of
  the tachyonic instability in Horndeski gravity},''
  \href{http://dx.doi.org/10.1088/1475-7516/2019/02/029}{{\em JCAP} {\bf 1902}
  (2019) no.~02, 029},
\href{http://arxiv.org/abs/1810.03461}{{\tt arXiv:1810.03461 [gr-qc]}}.

\bibitem{Horndeski:1974wa}
G.~W. Horndeski, ``{Second-order scalar-tensor field equations in a
  four-dimensional space},''
\href{http://dx.doi.org/10.1007/BF01807638}{{\em Int. J. Theor. Phys.} {\bf 10}
  (1974)  363--384}.

\bibitem{Deffayet:2011gz}
C.~Deffayet, X.~Gao, D.~A. Steer, and G.~Zahariade, ``{From k-essence to
  generalised Galileons},''
  \href{http://dx.doi.org/10.1103/PhysRevD.84.064039}{{\em Phys. Rev.} {\bf
  D84} (2011)  064039},
\href{http://arxiv.org/abs/1103.3260}{{\tt arXiv:1103.3260 [hep-th]}}.

\bibitem{Kobayashi:2011nu}
T.~Kobayashi, M.~Yamaguchi, and J.~Yokoyama, ``{Generalized G-inflation:
  Inflation with the most general second-order field equations},''
  \href{http://dx.doi.org/10.1143/PTP.126.511}{{\em Prog. Theor. Phys.} {\bf
  126} (2011)  511--529},
\href{http://arxiv.org/abs/1105.5723}{{\tt arXiv:1105.5723 [hep-th]}}.

\bibitem{Creminelli:2017sry}
P.~Creminelli and F.~Vernizzi, ``{Dark Energy after GW170817 and GRB170817A},''
  \href{http://dx.doi.org/10.1103/PhysRevLett.119.251302}{{\em Phys. Rev.
  Lett.} {\bf 119} (2017) no.~25, 251302},
\href{http://arxiv.org/abs/1710.05877}{{\tt arXiv:1710.05877 [astro-ph.CO]}}.

\bibitem{Sakstein:2017xjx}
J.~Sakstein and B.~Jain, ``{Implications of the Neutron Star Merger GW170817
  for Cosmological Scalar-Tensor Theories},''
  \href{http://dx.doi.org/10.1103/PhysRevLett.119.251303}{{\em Phys. Rev.
  Lett.} {\bf 119} (2017) no.~25, 251303},
\href{http://arxiv.org/abs/1710.05893}{{\tt arXiv:1710.05893 [astro-ph.CO]}}.

\bibitem{Ezquiaga:2017ekz}
J.~M. Ezquiaga and M.~Zumalacarregui, ``{Dark Energy After GW170817: Dead Ends
  and the Road Ahead},''
  \href{http://dx.doi.org/10.1103/PhysRevLett.119.251304}{{\em Phys. Rev.
  Lett.} {\bf 119} (2017) no.~25, 251304},
\href{http://arxiv.org/abs/1710.05901}{{\tt arXiv:1710.05901 [astro-ph.CO]}}.

\bibitem{Baker:2017hug}
T.~Baker, E.~Bellini, P.~G. Ferreira, M.~Lagos, J.~Noller, and I.~Sawicki,
  ``{Strong constraints on cosmological gravity from GW170817 and GRB
  170817A},'' \href{http://dx.doi.org/10.1103/PhysRevLett.119.251301}{{\em
  Phys. Rev. Lett.} {\bf 119} (2017) no.~25, 251301},
\href{http://arxiv.org/abs/1710.06394}{{\tt arXiv:1710.06394 [astro-ph.CO]}}.

\bibitem{Amendola:2012ky}
L.~Amendola, M.~Kunz, M.~Motta, I.~D. Saltas, and I.~Sawicki, ``{Observables
  and unobservables in dark energy cosmologies},''
  \href{http://dx.doi.org/10.1103/PhysRevD.87.023501}{{\em Phys. Rev.} {\bf
  D87} (2013) no.~2, 023501},
\href{http://arxiv.org/abs/1210.0439}{{\tt arXiv:1210.0439 [astro-ph.CO]}}.

\bibitem{Amendola:2014wma}
L.~Amendola, G.~Ballesteros, and V.~Pettorino, ``{Effects of modified gravity
  on B-mode polarization},''
  \href{http://dx.doi.org/10.1103/PhysRevD.90.043009}{{\em Phys. Rev.} {\bf
  D90} (2014)  043009},
\href{http://arxiv.org/abs/1405.7004}{{\tt arXiv:1405.7004 [astro-ph.CO]}}.

\bibitem{Deffayet:2010qz}
C.~Deffayet, O.~Pujolas, I.~Sawicki, and A.~Vikman, ``{Imperfect Dark Energy
  from Kinetic Gravity Braiding},''
  \href{http://dx.doi.org/10.1088/1475-7516/2010/10/026}{{\em JCAP} {\bf 1010}
  (2010)  026},
\href{http://arxiv.org/abs/1008.0048}{{\tt arXiv:1008.0048 [hep-th]}}.

\bibitem{Linder:2014fna}
E.~V. Linder, ``{Are Scalar and Tensor Deviations Related in Modified
  Gravity?},'' \href{http://dx.doi.org/10.1103/PhysRevD.90.083536}{{\em Phys.
  Rev.} {\bf D90} (2014) no.~8, 083536},
\href{http://arxiv.org/abs/1407.8184}{{\tt arXiv:1407.8184 [astro-ph.CO]}}.

\bibitem{Raveri:2014eea}
M.~Raveri, C.~Baccigalupi, A.~Silvestri, and S.-Y. Zhou, ``{Measuring the speed
  of cosmological gravitational waves},''
  \href{http://dx.doi.org/10.1103/PhysRevD.91.061501}{{\em Phys. Rev.} {\bf
  D91} (2015) no.~6, 061501(R)},
\href{http://arxiv.org/abs/1405.7974}{{\tt arXiv:1405.7974 [astro-ph.CO]}}.

\bibitem{Saltas:2014dha}
I.~D. Saltas, I.~Sawicki, L.~Amendola, and M.~Kunz, ``{Anisotropic Stress as a
  Signature of Nonstandard Propagation of Gravitational Waves},''
  \href{http://dx.doi.org/10.1103/PhysRevLett.113.191101}{{\em Phys. Rev.
  Lett.} {\bf 113} (2014) no.~19, 191101},
\href{http://arxiv.org/abs/1406.7139}{{\tt arXiv:1406.7139 [astro-ph.CO]}}.

\bibitem{Lombriser:2016yzn}
L.~Lombriser and N.~A. Lima, ``{Challenges to Self-Acceleration in Modified
  Gravity from Gravitational Waves and Large-Scale Structure},''
  \href{http://dx.doi.org/10.1016/j.physletb.2016.12.048}{{\em Phys. Lett.}
  {\bf B765} (2017)  382--385},
\href{http://arxiv.org/abs/1602.07670}{{\tt arXiv:1602.07670 [astro-ph.CO]}}.

\bibitem{Lombriser:2015sxa}
L.~Lombriser and A.~Taylor, ``{Breaking a Dark Degeneracy with Gravitational
  Waves},'' \href{http://dx.doi.org/10.1088/1475-7516/2016/03/031}{{\em JCAP}
  {\bf 1603} (2016) no.~03, 031},
\href{http://arxiv.org/abs/1509.08458}{{\tt arXiv:1509.08458 [astro-ph.CO]}}.

\bibitem{Jimenez:2015bwa}
J.~Beltran~Jimenez, F.~Piazza, and H.~Velten, ``{Evading the Vainshtein
  Mechanism with Anomalous Gravitational Wave Speed: Constraints on Modified
  Gravity from Binary Pulsars},''
  \href{http://dx.doi.org/10.1103/PhysRevLett.116.061101}{{\em Phys. Rev.
  Lett.} {\bf 116} (2016) no.~6, 061101},
\href{http://arxiv.org/abs/1507.05047}{{\tt arXiv:1507.05047 [gr-qc]}}.

\bibitem{Bettoni:2016mij}
D.~Bettoni, J.~M. Ezquiaga, K.~Hinterbichler, and M.~Zumalacarregui, ``{Speed
  of Gravitational Waves and the Fate of Scalar-Tensor Gravity},''
  \href{http://dx.doi.org/10.1103/PhysRevD.95.084029}{{\em Phys. Rev.} {\bf
  D95} (2017) no.~8, 084029},
\href{http://arxiv.org/abs/1608.01982}{{\tt arXiv:1608.01982 [gr-qc]}}.

\bibitem{Sawicki:2016klv}
I.~Sawicki, I.~D. Saltas, M.~Motta, L.~Amendola, and M.~Kunz, ``{Nonstandard
  gravitational waves imply gravitational slip: On the difficulty of partially
  hiding new gravitational degrees of freedom},''
  \href{http://dx.doi.org/10.1103/PhysRevD.95.083520}{{\em Phys. Rev.} {\bf
  D95} (2017) no.~8, 083520},
\href{http://arxiv.org/abs/1612.02002}{{\tt arXiv:1612.02002 [astro-ph.CO]}}.

\bibitem{Bellini:2014fua}
E.~Bellini and I.~Sawicki, ``{Maximal freedom at minimum cost: linear
  large-scale structure in general modifications of gravity},''
  \href{http://dx.doi.org/10.1088/1475-7516/2014/07/050}{{\em JCAP} {\bf 1407}
  (2014)  050},
\href{http://arxiv.org/abs/1404.3713}{{\tt arXiv:1404.3713 [astro-ph.CO]}}.

\bibitem{Linder:2015rcz}
E.~V. Linder, G.~Sengör, and S.~Watson, ``{Is the Effective Field Theory of
  Dark Energy Effective?},''
  \href{http://dx.doi.org/10.1088/1475-7516/2016/05/053}{{\em JCAP} {\bf 1605}
  (2016) no.~05, 053},
\href{http://arxiv.org/abs/1512.06180}{{\tt arXiv:1512.06180 [astro-ph.CO]}}.

\bibitem{Linder:2016wqw}
E.~V. Linder, ``{Challenges in connecting modified gravity theory and
  observations},'' \href{http://dx.doi.org/10.1103/PhysRevD.95.023518}{{\em
  Phys. Rev.} {\bf D95} (2017) no.~2, 023518},
\href{http://arxiv.org/abs/1607.03113}{{\tt arXiv:1607.03113 [astro-ph.CO]}}.

\bibitem{Denissenya:2018mqs}
M.~Denissenya and E.~V. Linder, ``{Gravity's Islands: Parametrizing Horndeski
  Stability},'' \href{http://dx.doi.org/10.1088/1475-7516/2018/11/010}{{\em
  JCAP} {\bf 1811} (2018) no.~11, 010},
\href{http://arxiv.org/abs/1808.00013}{{\tt arXiv:1808.00013 [astro-ph.CO]}}.

\bibitem{Lombriser:2018olq}
L.~Lombriser, C.~Dalang, J.~Kennedy, and A.~Taylor, ``{Inherently stable
  effective field theory for dark energy and modified gravity},''
  \href{http://dx.doi.org/10.1088/1475-7516/2019/01/041}{{\em JCAP} {\bf 1901}
  (2019) no.~01, 041},
\href{http://arxiv.org/abs/1810.05225}{{\tt arXiv:1810.05225 [astro-ph.CO]}}.

\bibitem{Gleyzes:2017kpi}
J.~Gleyzes, ``{Parametrizing modified gravity for cosmological surveys},''
  \href{http://dx.doi.org/10.1103/PhysRevD.96.063516}{{\em Phys. Rev.} {\bf
  D96} (2017) no.~6, 063516},
\href{http://arxiv.org/abs/1705.04714}{{\tt arXiv:1705.04714 [astro-ph.CO]}}.

\bibitem{Traykova:2021hbr}
D.~Traykova, E.~Bellini, P.~G. Ferreira, C.~Garc\'\i{}a-Garc\'\i{}a, J.~Noller,
  and M.~Zumalac\'arregui, ``{Theoretical priors in scalar-tensor cosmologies:
  Shift-symmetric Horndeski models},''
  \href{http://dx.doi.org/10.1103/PhysRevD.104.083502}{{\em Phys. Rev. D} {\bf
  104} (2021) no.~8, 083502}, \href{http://arxiv.org/abs/2103.11195}{{\tt
  arXiv:2103.11195 [astro-ph.CO]}}.

\bibitem{Ade:2015xua}
{\bf Planck} Collaboration, P.~Collaboration {\em et al.}, ``{Planck 2015
  results. XIII. Cosmological parameters},''
\href{http://arxiv.org/abs/1502.01589}{{\tt arXiv:1502.01589 [astro-ph.CO]}}.

\bibitem{Planck-Collaboration:2016aa}
{Planck Collaboration},
  \href{http://dx.doi.org/10.1051/0004-6361/201525941}{``{Planck 2015 results.
  XV. Gravitational lensing},''{\em Astronomy and Astrophysics} {\bf 594}
  (Sept., 2016)  A15}, \href{http://arxiv.org/abs/1502.01591}{{\tt
  arXiv:1502.01591}}.

\bibitem{Planck-Collaboration:2016af}
{Planck Collaboration},
  \href{http://dx.doi.org/10.1051/0004-6361/201526926}{``{Planck 2015 results.
  XI. CMB power spectra, likelihoods, and robustness of parameters},''{\em
  Astronomy and Astrophysics} {\bf 594} (Sept., 2016)  A11},
  \href{http://arxiv.org/abs/1507.02704}{{\tt arXiv:1507.02704}}.

\bibitem{Planck-Collaboration:2016ae}
{Planck Collaboration},
  \href{http://dx.doi.org/10.1051/0004-6361/201525830}{``{Planck 2015 results.
  XIII. Cosmological parameters},''{\em Astronomy and Astrophysics} {\bf 594}
  (Sept., 2016)  A13}, \href{http://arxiv.org/abs/1502.01589}{{\tt
  arXiv:1502.01589}}.

\bibitem{Planck:2019nip}
{\bf Planck} Collaboration, N.~Aghanim {\em et al.}, ``{Planck 2018 results. V.
  CMB power spectra and likelihoods},''
  \href{http://dx.doi.org/10.1051/0004-6361/201936386}{{\em Astron. Astrophys.}
  {\bf 641} (2020)  A5}, \href{http://arxiv.org/abs/1907.12875}{{\tt
  arXiv:1907.12875 [astro-ph.CO]}}.

\bibitem{Anderson:2014}
L.~{Anderson et al.}, \href{http://dx.doi.org/10.1093/mnras/stu523}{``{The
  clustering of galaxies in the SDSS-III Baryon Oscillation Spectroscopic
  Survey: baryon acoustic oscillations in the Data Releases 10 and 11 Galaxy
  samples},''{\em \mnras} {\bf 441} (June, 2014)  24--62},
  \href{http://arxiv.org/abs/1312.4877}{{\tt arXiv:1312.4877}}.

\bibitem{Ross:2015}
A.~J. {Ross}, L.~{Samushia}, C.~{Howlett}, W.~J. {Percival}, A.~{Burden}, and
  M.~{Manera}, \href{http://dx.doi.org/10.1093/mnras/stv154}{``{The clustering
  of the SDSS DR7 main Galaxy sample - I. A 4 per cent distance measure at z =
  0.15},''{\em \mnras} {\bf 449} (May, 2015)  835--847},
  \href{http://arxiv.org/abs/1409.3242}{{\tt arXiv:1409.3242}}.

\bibitem{Tegmark:2006}
M.~{Tegmark et al.},
  \href{http://dx.doi.org/10.1103/PhysRevD.74.123507}{``{Cosmological
  constraints from the SDSS luminous red galaxies},''{\em \prd} {\bf 74} (Dec.,
  2006)  123507}, \href{http://arxiv.org/abs/astro-ph/0608632}{{\tt
  astro-ph/0608632}}.

\bibitem{Beutler:2012}
F.~{Beutler}, C.~{Blake}, M.~{Colless}, D.~H. {Jones}, L.~{Staveley-Smith},
  G.~B. {Poole}, L.~{Campbell}, Q.~{Parker}, W.~{Saunders}, and F.~{Watson},
  \href{http://dx.doi.org/10.1111/j.1365-2966.2012.21136.x}{``{The 6dF Galaxy
  Survey: $z \approx 0$ measurements of the growth rate and
  {$\sigma$}$_{8}$},''{\em \mnras} {\bf 423} (July, 2012)  3430--3444},
  \href{http://arxiv.org/abs/1204.4725}{{\tt arXiv:1204.4725}}.

\bibitem{Samushia:2014}
L.~{Samushia et al.}, \href{http://dx.doi.org/10.1093/mnras/stu197}{``{The
  clustering of galaxies in the SDSS-III Baryon Oscillation Spectroscopic
  Survey: measuring growth rate and geometry with anisotropic
  clustering},''{\em \mnras} {\bf 439} (Apr., 2014)  3504--3519},
  \href{http://arxiv.org/abs/1312.4899}{{\tt arXiv:1312.4899}}.

\bibitem{Gumrukcuoglu:2016jbh}
A.~E. Gumrukcuoglu, S.~Mukohyama, and T.~P. Sotiriou, ``{Low energy ghosts and
  the Jeans’ instability},''
  \href{http://dx.doi.org/10.1103/PhysRevD.94.064001}{{\em Phys. Rev.} {\bf
  D94} (2016) no.~6, 064001},
\href{http://arxiv.org/abs/1606.00618}{{\tt arXiv:1606.00618 [hep-th]}}.

\bibitem{Wolf:2019hzy}
W.~J. Wolf and M.~Lagos, ``{Cosmological Instabilities and the Role of Matter
  Interactions in Dynamical Dark Energy Models},''
  \href{http://dx.doi.org/10.1103/PhysRevD.100.084035}{{\em Phys. Rev. D} {\bf
  100} (2019) no.~8, 084035}, \href{http://arxiv.org/abs/1908.03212}{{\tt
  arXiv:1908.03212 [gr-qc]}}.

\bibitem{1902RSPTA.199....1J}
J.~H. {Jeans}, \href{http://dx.doi.org/10.1098/rsta.1902.0012}{``The stability
  of a spherical nebula,''{\em Phil. Trans. R. Soc. A} {\bf 199} (Jan., 1902)
  1--53}.

\bibitem{Frusciante:2019puu}
N.~Frusciante, S.~Peirone, L.~Atayde, and A.~De~Felice, ``{Phenomenology of the
  generalized cubic covariant Galileon model and cosmological bounds},''
  \href{http://dx.doi.org/10.1103/PhysRevD.101.064001}{{\em Phys. Rev.} {\bf
  D101} (2020) no.~6, 064001},
\href{http://arxiv.org/abs/1912.07586}{{\tt arXiv:1912.07586 [astro-ph.CO]}}.

\bibitem{DeFelice:2011bh}
A.~De~Felice and S.~Tsujikawa, ``{Conditions for the cosmological viability of
  the most general scalar-tensor theories and their applications to extended
  Galileon dark energy models},''
  \href{http://dx.doi.org/10.1088/1475-7516/2012/02/007}{{\em JCAP} {\bf 02}
  (2012)  007}, \href{http://arxiv.org/abs/1110.3878}{{\tt arXiv:1110.3878
  [gr-qc]}}.

\bibitem{Arroja_2010}
F.~Arroja and M.~Sasaki, ``{A note on the equivalence of a barotropic perfect
  fluid with a K-essence scalar field},''
  \href{http://dx.doi.org/10.1103/PhysRevD.81.107301}{{\em Phys. Rev. D} {\bf
  81} (2010)  107301}, \href{http://arxiv.org/abs/1002.1376}{{\tt
  arXiv:1002.1376 [astro-ph.CO]}}.

\bibitem{Christopherson_2009}
A.~J. Christopherson and K.~A. Malik,
  \href{http://dx.doi.org/10.1016/j.physletb.2009.04.003}{``The non-adiabatic
  pressure in general scalar field systems,''{\em Physics Letters B} {\bf 675}
  (May, 2009)  159–163}.
  \url{http://dx.doi.org/10.1016/j.physletb.2009.04.003}.

\bibitem{Faraoni_2012}
V.~Faraoni, ``{The correspondence between a scalar field and an effective
  perfect fluid},'' \href{http://dx.doi.org/10.1103/PhysRevD.85.024040}{{\em
  Phys. Rev. D} {\bf 85} (2012)  024040},
  \href{http://arxiv.org/abs/1201.1448}{{\tt arXiv:1201.1448 [gr-qc]}}.

\bibitem{Boubekeur:2008kn}
L.~Boubekeur, P.~Creminelli, J.~Norena, and F.~Vernizzi, ``{Action approach to
  cosmological perturbations: the 2nd order metric in matter dominance},''
  \href{http://dx.doi.org/10.1088/1475-7516/2008/08/028}{{\em JCAP} {\bf 08}
  (2008)  028}, \href{http://arxiv.org/abs/0806.1016}{{\tt arXiv:0806.1016
  [astro-ph]}}.

\bibitem{DeFelice:2015moy}
A.~De~Felice and S.~Mukohyama, ``{Phenomenology in minimal theory of massive
  gravity},'' \href{http://dx.doi.org/10.1088/1475-7516/2016/04/028}{{\em JCAP}
  {\bf 04} (2016)  028}, \href{http://arxiv.org/abs/1512.04008}{{\tt
  arXiv:1512.04008 [hep-th]}}.

\bibitem{Schutz:1977df}
B.~F. Schutz and R.~Sorkin, ``{Variational aspects of relativistic field
  theories, with application to perfect fluids},''
  \href{http://dx.doi.org/10.1016/0003-4916(77)90200-7}{{\em Annals Phys.} {\bf
  107} (1977)  1--43}.

\bibitem{Brown:1992kc}
J.~D. Brown, ``{Action functionals for relativistic perfect fluids},''
  \href{http://dx.doi.org/10.1088/0264-9381/10/8/017}{{\em Class. Quant. Grav.}
  {\bf 10} (1993)  1579--1606}, \href{http://arxiv.org/abs/gr-qc/9304026}{{\tt
  arXiv:gr-qc/9304026}}.

\bibitem{DIEZ_TEJEDOR_2005}
A.~DIEZ-TEJEDOR and A.~FEINSTEIN,
  \href{http://dx.doi.org/10.1142/s0218271805007152}{``Relativistic
  hydrodynamics with sources for cosmological k-fluids,''{\em International
  Journal of Modern Physics D} {\bf 14} (Sep, 2005)  1561–1576}.
  \url{http://dx.doi.org/10.1142/S0218271805007152}.

\bibitem{PhysRevD.96.024060}
A.~De~Felice, N.~Frusciante, and G.~Papadomanolakis,
  \href{http://dx.doi.org/10.1103/PhysRevD.96.024060}{``de sitter limit
  analysis for dark energy and modified gravity models,''{\em Phys. Rev. D}
  {\bf 96} (Jul, 2017)  024060}.
  \url{https://link.aps.org/doi/10.1103/PhysRevD.96.024060}.

\bibitem{Bellini:2019syt}
E.~Bellini, I.~Sawicki, and M.~Zumalac\'arregui, ``{hi\_class: Background
  Evolution, Initial Conditions and Approximation Schemes},''
  \href{http://dx.doi.org/10.1088/1475-7516/2020/02/008}{{\em JCAP} {\bf 02}
  (2020)  008}, \href{http://arxiv.org/abs/1909.01828}{{\tt arXiv:1909.01828
  [astro-ph.CO]}}.

\bibitem{Creminelli:2019kjy}
P.~Creminelli, G.~Tambalo, F.~Vernizzi, and V.~Yingcharoenrat, ``{Dark-Energy
  Instabilities induced by Gravitational Waves},''
\href{http://arxiv.org/abs/1910.14035}{{\tt arXiv:1910.14035 [gr-qc]}}.

\bibitem{Creminelli:2019nok}
P.~Creminelli, G.~Tambalo, F.~Vernizzi, and V.~Yingcharoenrat, ``{Resonant
  Decay of Gravitational Waves into Dark Energy},''
  \href{http://dx.doi.org/10.1088/1475-7516/2019/10/072}{{\em JCAP} {\bf 1910}
  (2019) no.~10, 072},
\href{http://arxiv.org/abs/1906.07015}{{\tt arXiv:1906.07015 [gr-qc]}}.

\bibitem{Creminelli:2018xsv}
P.~Creminelli, M.~Lewandowski, G.~Tambalo, and F.~Vernizzi, ``{Gravitational
  Wave Decay into Dark Energy},''
  \href{http://dx.doi.org/10.1088/1475-7516/2018/12/025}{{\em JCAP} {\bf 1812}
  (2018) no.~12, 025},
\href{http://arxiv.org/abs/1809.03484}{{\tt arXiv:1809.03484 [astro-ph.CO]}}.

\bibitem{Babichev:2011iz}
E.~Babichev, C.~Deffayet, and G.~Esposito-Farese, ``{Constraints on
  Shift-Symmetric Scalar-Tensor Theories with a Vainshtein Mechanism from
  Bounds on the Time Variation of G},''
  \href{http://dx.doi.org/10.1103/PhysRevLett.107.251102}{{\em Phys. Rev.
  Lett.} {\bf 107} (2011)  251102}, \href{http://arxiv.org/abs/1107.1569}{{\tt
  arXiv:1107.1569 [gr-qc]}}.

\bibitem{Burrage:2020jkj}
C.~Burrage and J.~Dombrowski, ``{Constraining the cosmological evolution of
  scalar-tensor theories with local measurements of the time variation of G},''
  \href{http://arxiv.org/abs/2004.14260}{{\tt arXiv:2004.14260 [astro-ph.CO]}}.

\bibitem{Heisenberg:2020cyi}
L.~Heisenberg, J.~Noller, and J.~Zosso, ``{Horndeski under the quantum
  loupe},'' \href{http://dx.doi.org/10.1088/1475-7516/2020/10/010}{{\em JCAP}
  {\bf 10} (2020)  010}, \href{http://arxiv.org/abs/2004.11655}{{\tt
  arXiv:2004.11655 [hep-th]}}.

\bibitem{Pirtskhalava:2015nla}
D.~Pirtskhalava, L.~Santoni, E.~Trincherini, and F.~Vernizzi, ``{Weakly Broken
  Galileon Symmetry},''
  \href{http://dx.doi.org/10.1088/1475-7516/2015/09/007}{{\em JCAP} {\bf 1509}
  (2015) no.~09, 007},
\href{http://arxiv.org/abs/1505.00007}{{\tt arXiv:1505.00007 [hep-th]}}.

\bibitem{Melville:2019wyy}
S.~Melville and J.~Noller, ``{Positivity in the Sky: Constraining dark energy
  and modified gravity from the UV},''
  \href{http://dx.doi.org/10.1103/PhysRevD.101.021502}{{\em Phys. Rev. D} {\bf
  101} (2020) no.~2, 021502}, \href{http://arxiv.org/abs/1904.05874}{{\tt
  arXiv:1904.05874 [astro-ph.CO]}}. [Erratum: Phys.Rev.D 102, 049902 (2020)].

\bibitem{Kennedy:2020ehn}
J.~Kennedy and L.~Lombriser, ``{Positivity bounds on reconstructed Horndeski
  models},''
\href{http://arxiv.org/abs/2003.05318}{{\tt arXiv:2003.05318 [gr-qc]}}.

\bibitem{deRham:2021fpu}
C.~de~Rham, S.~Melville, and J.~Noller, ``{Positivity Bounds on Dark Energy:
  When Matter Matters},'' \href{http://arxiv.org/abs/2103.06855}{{\tt
  arXiv:2103.06855 [astro-ph.CO]}}.

\bibitem{Blas:2011rf}
D.~Blas, J.~Lesgourgues, and T.~Tram, ``{The Cosmic Linear Anisotropy Solving
  System (CLASS) II: Approximation schemes},''
  \href{http://dx.doi.org/10.1088/1475-7516/2011/07/034}{{\em JCAP} {\bf 1107}
  (2011)  034},
\href{http://arxiv.org/abs/1104.2933}{{\tt arXiv:1104.2933 [astro-ph.CO]}}.

\bibitem{corner}
D.~Foreman-Mackey, ``corner.py: Scatterplot matrices in python,''
  \href{http://dx.doi.org/10.21105/joss.00024}{{\em The Journal of Open Source
  Software} {\bf 24} (2016)  }. \url{http://dx.doi.org/10.5281/zenodo.45906}.

\bibitem{Audren:2012wb}
B.~Audren, J.~Lesgourgues, K.~Benabed, and S.~Prunet, ``{Conservative
  Constraints on Early Cosmology: an illustration of the Monte Python
  cosmological parameter inference code},''
  \href{http://dx.doi.org/10.1088/1475-7516/2013/02/001}{{\em JCAP} {\bf 1302}
  (2013)  001},
\href{http://arxiv.org/abs/1210.7183}{{\tt arXiv:1210.7183 [astro-ph.CO]}}.

\bibitem{Brinckmann:2018cvx}
T.~Brinckmann and J.~Lesgourgues, ``{MontePython 3: boosted MCMC sampler and
  other features},''
\href{http://arxiv.org/abs/1804.07261}{{\tt arXiv:1804.07261 [astro-ph.CO]}}.

\bibitem{xAct}
J.~M. Mart\'in-Garc\'ia, ``{xAct 2002-2014},'' {\em http://www.xact.es/}  .

\end{thebibliography}\endgroup
\end{document}